\documentclass[nobibnotes, twocolumn, aps, prb, superscriptaddress]{revtex4-1}
\usepackage[utf8]{inputenc}
\usepackage[letterpaper, margin=1.8cm]{geometry}
\usepackage{graphicx}
\usepackage{dcolumn}
\usepackage{bm}
\usepackage{amsmath}
\usepackage[english]{babel}
\usepackage[T1]{fontenc}
\usepackage{verbatim}
\usepackage{physics}
\usepackage{booktabs}
\usepackage{gensymb}
\usepackage{amssymb}
\usepackage{placeins}
\usepackage{tikz}
\usepackage{xcolor}

\usetikzlibrary{shapes}

\definecolor{color1}{HTML}{1f77b4}
\definecolor{color2}{HTML}{ff7f0e}
\definecolor{color3}{HTML}{2ca02c}
\definecolor{color4}{HTML}{d62728}
\definecolor{color5}{HTML}{9467bd}
\definecolor{color6}{HTML}{8c564b}
\definecolor{color7}{HTML}{e377c2}
\definecolor{color8}{HTML}{7f7f7f}

\DeclareMathOperator{\diag}{diag}
\usepackage{textcomp}

\makeatletter
\newcommand\setcurrentname[1]{\def\@currentlabelname{#1}}
\def\maketitle{
\@author@finish
\title@column\titleblock@produce
\suppressfloats[t]}
\makeatother

\DeclareUnicodeCharacter{03C0}{$\pi$}
\DeclareUnicodeCharacter{03BC}{$\mu$}

\begin{document}

\title{Sweet-spot operation of a germanium hole spin qubit with highly anisotropic noise sensitivity}

\author{N.W.~Hendrickx}
\email{nwh@zurich.ibm.com}
\author{L. Massai}
\author{M. Mergenthaler}
\author{F. Schupp}
\author{S. Paredes}
\affiliation{IBM Research Europe~-~Zurich, Säumerstrasse 4, 8803 Rüschlikon, Switzerland}
\author{S.W. Bedell}
\affiliation{IBM Quantum, T.J. Watson Research Center, 1101 Kitchawan Road, Yorktown Heights, New York 10598, USA}
\author{G. Salis}
\author{A. Fuhrer}
\email{afu@zurich.ibm.com}
\affiliation{IBM Research Europe~-~Zurich, Säumerstrasse 4, 8803 Rüschlikon, Switzerland}
\date{\today}

\begin{abstract}
Spin qubits defined by valence band hole states comprise an attractive candidate for quantum information processing due to their inherent coupling to electric fields enabling fast and scalable qubit control. In particular, heavy holes in germanium have shown great promise, with recent demonstrations of fast and high-fidelity qubit operations. However, the mechanisms and anisotropies that underlie qubit driving and decoherence are still mostly unclear. Here, we report on the highly anisotropic heavy-hole $g$-tensor and its dependence on electric fields, allowing us to relate both qubit driving and decoherence to an electric modulation of the $g$-tensor. We also confirm the predicted Ising-type hyperfine interaction but show that qubit coherence is ultimately limited by $1/f$ charge noise. Finally, we operate the qubit at low magnetic field and measure a dephasing time of $T_2^*=9.2~\mu$s, while maintaining a single-qubit gate fidelity of 99.94~\%, that remains well above 99~\% at an operation temperature T>1~K. This understanding of qubit driving and decoherence mechanisms are key for the design and operation of scalable and highly coherent hole qubit arrays.
\end{abstract}
\maketitle

\section*{Introduction}
The development of a fault-tolerant quantum computer~\cite{knill_resilient_1998} able to solve relevant problems~\cite{reiher_elucidating_2017} requires the integration of many highly coherent qubits. Spin qubits based on quantum dots~\cite{loss_quantum_1998} hold excellent promise for scaling towards large-scale quantum processors, due to their small footprint and long coherence. Recently, great progress has been made, with demonstrations of single-qubit gate~\cite{veldhorst_addressable_2014, lawrie_simultaneous_2021}, two-qubit gate~\cite{mills_two-qubit_2022, xue_quantum_2022, tanttu_stability_2023}, and readout~\cite{harvey-collard_high-fidelity_2018} fidelities well above the fault-tolerant threshold. Furthermore, rudimentary quantum algorithms and simulations have been executed on multi-qubit arrays~\cite{hendrickx_four-qubit_2021, philips_universal_2022} including minimal error correction schemes~\cite{van_riggelen_phase_2022, takeda_quantum_2022} and compatibility with semiconductor manufacturing has been demonstrated~\cite{zwerver_qubits_2022}.

In particular, hole qubits in strained germanium quantum wells have gained a strong interest over recent years~\cite{scappucci_germanium_2020}, with demonstrations of single~\cite{hendrickx_fast_2020, jirovec_singlet-triplet_2021} and multi-qubit~\cite{hendrickx_four-qubit_2021, van_riggelen_phase_2022} operations and first steps towards the operation of large, multiplexed qubit registers~\cite{li_crossbar_2018, borsoi_shared_2022}. This surge of interest is rooted in the combination of favourable properties that holes in germanium possess: a low-effective mass that relaxes the constraints on device fabrication~\cite{sammak_shallow_2019} and a low-disorder quantum well that provides a low-noise qubit environment~\cite{lodari_low_2021} and enables excellent quantum dot control~\cite{lawrie_quantum_2020}, without the complication of low-energy valley states that have hindered progress for electrons in silicon.

The spin properties of valence band holes can be highly anisotropic~\cite{winkler_spin-orbit_2003, piot_single_2022, fischer_spin_2008, jirovec_dynamics_2022, zhang_anisotropic_2021}, which results in a field-dependent coupling to the two dominant sources of decoherence in spin qubits: nuclear spin fluctuations~\cite{veldhorst_addressable_2014} and charge noise~\cite{yoneda_quantum-dot_2018}. These anisotropies present both opportunities and challenges for building a scalable qubit platform. For example, the anisotropic heavy hole $g$-tensor can amplify small variations in quantum dot confinement, leading to site-dependent qubit properties~\cite{jirovec_dynamics_2022, hendrickx_four-qubit_2021} and increasing requirements on material uniformity. However, when well controlled, the anisotropy enables operational sweet spots where qubit control is maximized while decoherence is minimized~\cite{bosco_squeezed_2021, wang_modelling_2022, wang_optimal_2021, piot_single_2022}, overcoming the general trade-off between qubit control and coherence. Theoretical considerations predict the operating point of such sweet spots to depend on the exact material and device parameters like strain~\cite{abadillo-uriel_hole_2022} or device geometry~\cite{martinez_hole_2022}, but an experimental demonstration of the heavy hole anisotropies and their implications on qubit performance is lacking.

\begin{figure*}[htp]
\includegraphics{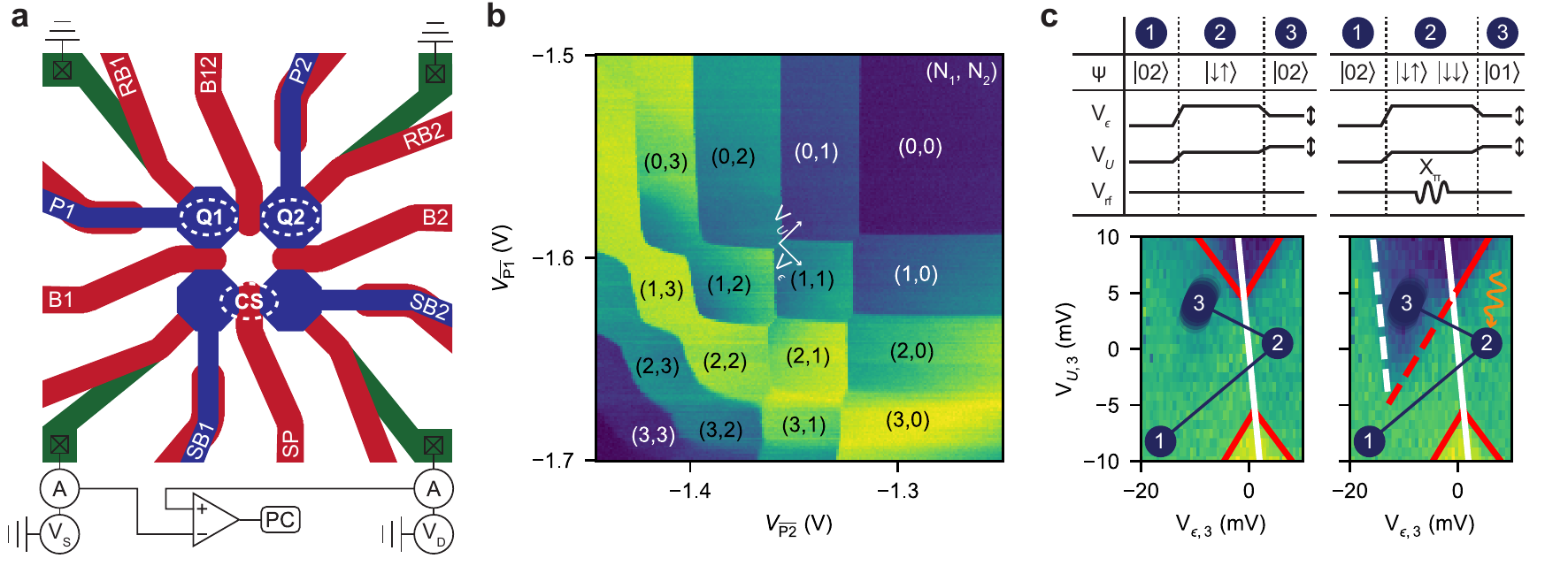}
\caption{\textbf{A germanium hole two-qubit system.} 
    \textbf{a}, Schematic drawing of the three quantum dot device. We define qubits Q1 and Q2 underneath plunger gates P1 and P2 respectively, that can be read out using the nearby charge sensor (CS) defined by gates SP, SB1, and SB2. The coupling between the qubits is controlled by B12 while the coupling of Q1 (Q2) to its reservoir is controlled by RB1 (RB2). We record the response of the charge sensor by measuring the differential current flowing into and out of the source (S) and drain (D) contacts.
    \textbf{b}, Two-quantum-dot charge stability diagram as a function of two virtualized plunger gate voltages $V_{\overline{\mathrm{P}_1}}$ and $V_{\overline{\mathrm{P}_2}}$. The different charge configurations are indicated by the numbers in parentheses ($N_1$, $N_2$). The direction of the virtual detuning $\epsilon$ and on-site energy $U$ axes are indicated.
    \textbf{c}, Spin-to-charge conversion is performed by latched Pauli spin blockade readout. The pulses applied to the $\epsilon$ and $U$ axes, as well as the qubit drive pulses $V_\text{rf}$ are shown in the top panels. The spins are initialized in the $\ket{\downarrow\uparrow}$ state by adiabatically sweeping across the interdot transition (1$\rightarrow$2). Next we apply either no pulse (left panel) or a $X_\pi$ pulse (right panel) to Q2 (2) and sweep (2$\rightarrow$3) to the readout point ($V_{\epsilon,3}$, $V_{U,3}$), which is rasterized to compose the entire map. Red lines indicate (extended) lead transition lines, while the white lines corresponds to the interdot transition lines of the quantum dot ground (solid) and excited (dashed) states.
    }
\label{fig:dots}
\end{figure*}

\section*{Germanium two-qubit device}
Here, we unveil the mechanisms that enable qubit driving and mediate decoherence in germanium hole qubits. We fully characterize the heavy-hole $g$-tensors of a two-qubit system and their sensitivity to electric fields. A comparison with the dependence of qubit coherence and Rabi frequency on the orientation and magnitude of the external magnetic field demonstrates that both qubit driving and charge-noise induced qubit decoherence are explained by the distortion of the g-tensor through electric fields. Furthermore, we confirm the predicted Ising character of the hyperfine interaction between the heavy-hole spin and the $^{73}$Ge nuclear spin bath, leading to a strong suppression of hyperfine coupling when the magnetic field is oriented in the plane of the qubit $g$-tensor. This understanding enables us to find an optimal operation regime that yields an improvement in spin coherence times of more than an order of magnitude compared to the state-of-the-art.

We define a two-qubit system based on confined hole spins in a strained Ge/SiGe heterostructure quantum well~\cite{bedell_low-temperature_2020}. The spins are confined in gate-defined quantum dots, formed respectively underneath plunger gates P1 and P2, with an additional gate B12 controlling the interdot coupling (see Fig.~\ref{fig:dots}a). Additionally, we form a large quantum dot underneath gate SP to act as a charge sensor of which the tunnel rates to in-diffused PtSiGe ohmic leads can be controlled by gates SB1 and SB2. We measure the charge sensor conductance to detect nearby tunnelling events. Using two virtual gates $\overline{\text{P1}}$ and $\overline{\text{P2}}$ (see Methods), we measure the charge stability diagram as plotted in Fig.~\ref{fig:dots}b. Well-defined charge occupancy regions can be observed, with the top right corner corresponding to both dots being fully depleted. We operate the device in the (1,1) charge region and perform latched Pauli spin blockade readout~\cite{ono_current_2002, harvey-collard_high-fidelity_2018, hendrickx_four-qubit_2021}, as shown in Fig.~\ref{fig:dots}c, where a distinct difference in the differential charge sensor current can be observed for the preparation of a $\ket{\downarrow\downarrow}$ and $\ket{\downarrow\uparrow}$ state.

\begin{figure*}[htp]
\includegraphics{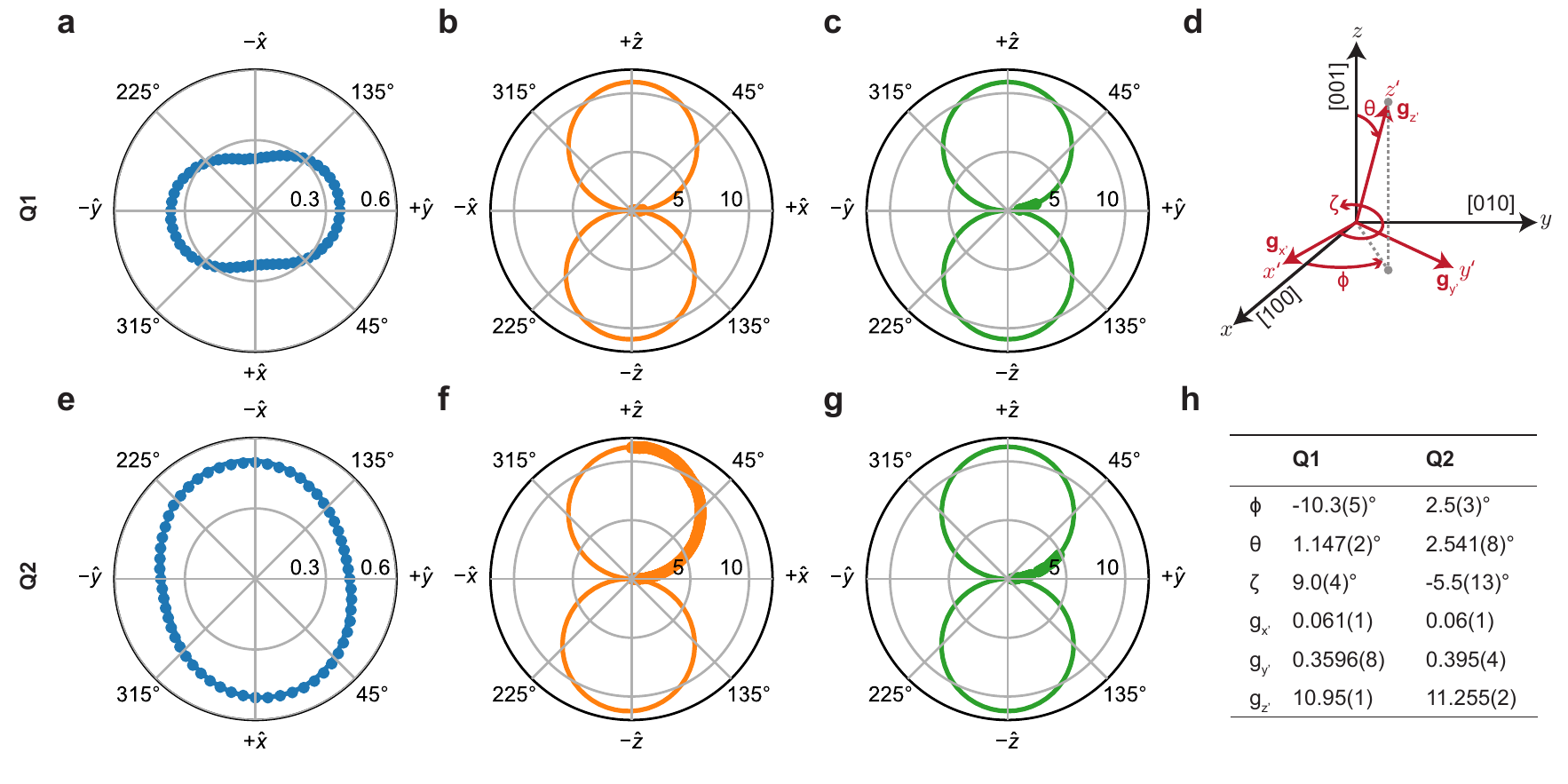}
\caption{\textbf{Measurement of the hole $\bm{g}$-tensor.} 
    \textbf{a}-\textbf{c}, \textbf{e}-\textbf{g}, Cross section of the $g$-tensor of Q1 (\textbf{a}-\textbf{c}) and Q2 (\textbf{e}-\textbf{g}) in the $xy$-plane (\textbf{a}, \textbf{d}), $xz$-plane (\textbf{b}, \textbf{e}), and the $yz$-plane (\textbf{c},  \textbf{f}) of the magnet frame. Dots indicate measurements of $g^*$ and the solid line corresponds to the fit of the $g$-tensor. Exemplary EDSR spectra used to extract $g$ are plotted in Supp.~Fig.~\ref{fig:edsr_spectra}.
    \textbf{d}, Diagram indicating the $zyz$-Euler rotation angles $\phi$, $\theta$, $\zeta$ of the principle $g$-tensor axes $\mathbf{g}_{x'}$, $\mathbf{g}_{y'}$, and $\mathbf{g}_{z'}$. The approximate crystal directions are indicated in brackets.
    \textbf{h}, Overview of the three $zyz$ Euler angles $\phi$, $\theta$, and $\zeta$ of the rotation of a $g$-tensor with principle components $g_{x'}$, $g_{y'}$, and $g_{z'}$, for Q1 and Q2.
    }
\label{fig:g_tensor}
\end{figure*}

\section*{Heavy hole $\bm{g}$-tensor}
The confinement of holes in a two-dimensional (2D) strained Ge quantum well splits heavy hole (HH) and light hole (LH) bands, with the former defining the ground state~\cite{sammak_shallow_2019}. As the electrical confinement in the plane of the quantum well is notably weaker than the  confinement in growth direction, the hole wave function is expected to contain mostly HH components~\cite{sammak_shallow_2019, wang_modelling_2022}. The degree of HH-LH mixing will affect the observed hole $g$-tensor, which is predicted to be highly anisotropic for the heavy hole states and much more isotropic for the light hole states~\cite{winkler_spin_2008}. The general $g$-tensor can be modelled as a rotated diagonal 3x3 matrix $\tensor{g}=R(\phi,~\theta,~\zeta)\diag{(g_{x'}, g_{y'}, g_{z'})}R^{-1}(\phi,~\theta,~\zeta)$, where $\phi$, $\theta$, and $\zeta$ are Euler angles corresponding to successive intrinsic rotations around axes $zyz$ and $g_{x'}$, $g_{y'}$, and $g_{z'}$ define the effective $g$-factors along the principle axes $x'y'z'$ of the $g$-tensor (Fig.~\ref{fig:g_tensor}d). We reconstruct the $g$-tensor for both Q1 and Q2 by measuring the effective $g$-factor $g^*=hf_{\text{Q}}/(\mu_\text{B}B)$, with $h$ the Planck constant, $f_{\text{Q}}=\abs{\mathbf{f}_{\text{Q}}}$ the qubit Larmor frequency, and $\mu_\text{B}$ the Bohr magneton, for different magnetic field orientations $\mathbf{B}=B\hat{\bm{b}}$. The measured data and fit of $\tensor{g}$ are plotted in Fig.~\ref{fig:g_tensor}a-c,e-g for cuts through the $xy$, $xz$, and $yz$ planes respectively. The observed $g$-tensor is found to be extremely anisotropic for both qubits, with $g_{z'}\approx30g_{y'}\approx180g_{x'}$, and the largest principle axis almost aligned to the sample growth direction $z$. The $g$-tensors of the two qubits are remarkably identical, with their principle axes lengths differing by $<10\%$, the azimuth rotations $\phi$ and $\zeta$ by $<15\degree$, and the elevation $\theta$ by less than $2\degree$ (see Fig.~\ref{fig:g_tensor}h). 

Due to the strong anisotropy, the qubit quantisation axis $h\mathbf{f}_\text{Q}=\mu_\text{B}\tensor{g}\mathbf{B}$ is not necessarily aligned with the applied magnetic field $\mathbf{B}$ as is the case in isotropic systems. In particular, any small deviation of $\mathbf{B}$ from the plane spanned by the two minor principal axes $x'y'$ of the $g$-tensor will result in a strong rotation of the qubit quantisation axis towards $\pm \hat{z}$ (see Supp. Fig.~\ref{fig:projection}b,c). Therefore, small variations between qubit $g$-tensors can still lead to a sizeable difference between the qubit quantisation axes, in particular for in-plane magnetic field orientations. Because Pauli spin blockade readout measures the relative spin projection of two qubits, we observe readout to be affected for these magnetic field orientations and being completely suppressed when the angle between $\mathbf{f_\text{Q1}}$ and $\mathbf{f_\text{Q2}}$ equals $\pi/2$ (Supp. Fig.~\ref{fig:projection}f,g). While the anisotropy between $g_{z'}$ and $g_{x',y'}$ is expected from the quantum well confinement, the in-plane anisotropy points to a non-circular confinement of the quantum dot. This can be explained by the device layout, as the finite potential on the interdot barrier breaks the symmetry of the individual quantum dots and gives rise to an non-circular confinement potential. We suspect that the small tilt of the $g$-tensor with respect to the sample axes is caused by localized strain gradients as imposed by the nanostructured gate electrodes~\cite{abadillo-uriel_hole_2022, corley-wiciak_nanoscale_2023}.

\begin{figure*}[htp]
\includegraphics{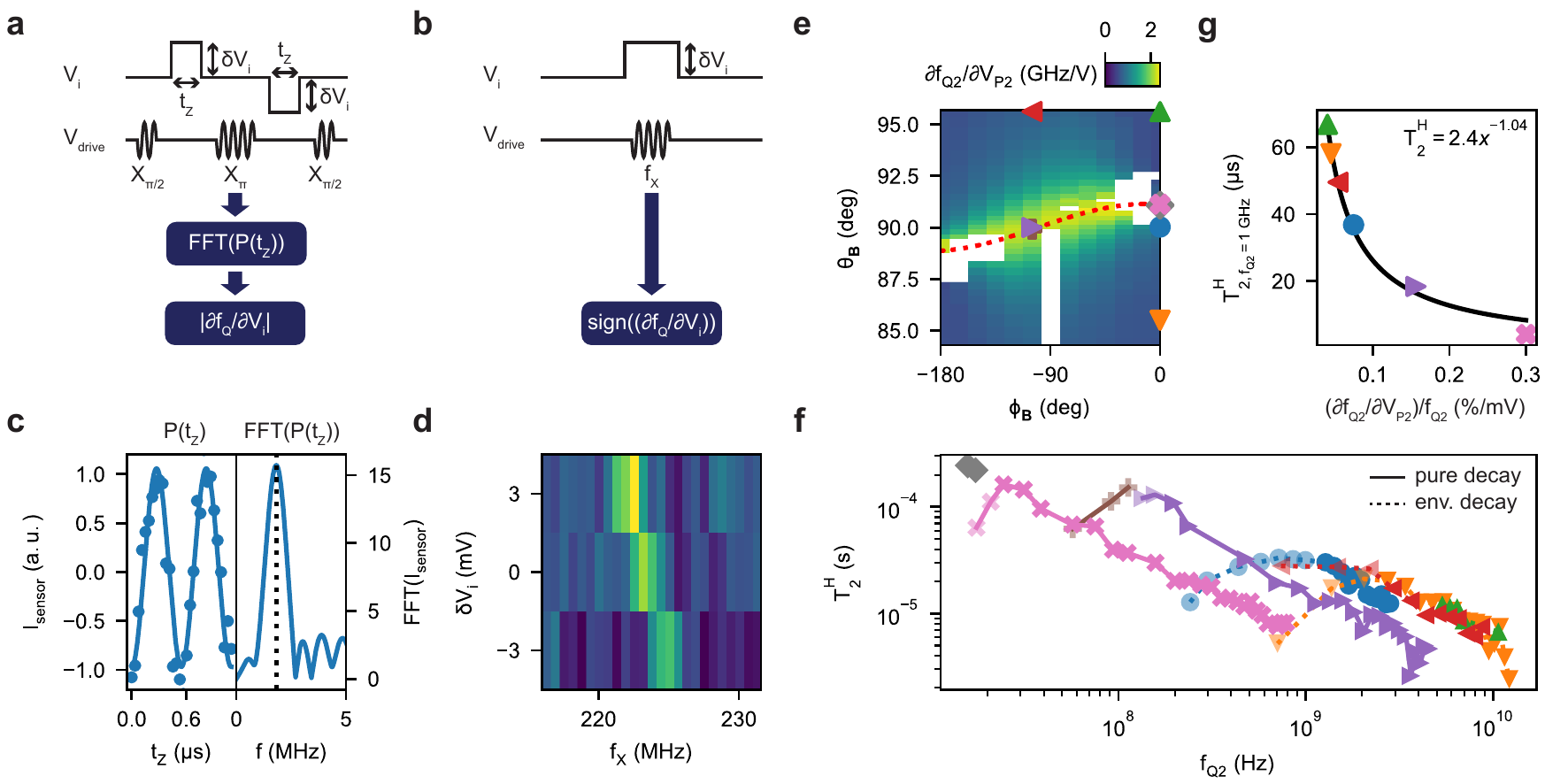}
\caption{\textbf{Electric field sensitivity and coherence dependence on magnetic field orientation.}
    \textbf{a}, Pulse sequences used to measure the voltage sensitivity of the energy splitting $\partial f_\text{Q}/\partial V_i$. A positive (negative) voltage pulse $\delta V_i$ of varying length $t_Z$ is applied to the test gate electrode $i$ in the first (second) free evolution of a Hahn echo to extract $|\partial f_\text{Q}/\partial V_i|$. 
    \textbf{b}, Pulse sequences used to infer the sign of $\partial f_\text{Q}/\partial V_i$ by assessing the shift of the qubit resonance frequency as a result of a voltage pulse $\delta V_i$.
    \textbf{c}, Left: charge sensor current $I_\text{sensor}$ as a function of $t_Z$, solid line is a fit to the data. Right: fast Fourier transform of $I_\text{sensor}$, allowing us to extract $|\partial f_\text{Q}/\partial V_i|$.
    \textbf{d}, $I_\text{sensor}$ as a function of the drive frequency $f_X$ and $\delta V_i$. The shift of the resonance frequency allows us to extract the sign of $\partial f_\text{Q}/\partial V_i$.
    \textbf{e}, The qubit energy splitting sensitivity to a voltage change on the plunger gate $\partial f_\text{Q2}/\partial V_\text{P2}$, as a function of different magnetic field orientations $\phi_\mathbf{B}$ and $\theta_\mathbf{B}$. $B$ is adapted to keep $f_\text{Q2}$ constant at $f_\text{Q2}=1.36(7)$~GHz. Data acquisition is hindered for the white areas as a result of limited qubit readout or addressability for these magnetic field orientations.
    \textbf{f}, Hahn coherence time $T_2^\text{H}$ as a function of the qubit frequency $f_\text{Q2}$, for different magnetic field orientations, indicated by the coloured markers in~\textbf{e} (exact field orientation in Supp.~Table~\ref{tab:field_angle}). Solid lines correspond to $T_2^\text{H}$ as extracted from a pure decay, while dashed lines correspond to $T_2^\text{H}$ as extracted from the envelope of the nuclear spin induced collapse-and-revival. Data indicated by opaque markers are used to fit the power law dependence of $T_2^\text{H}$.
    \textbf{g}, Expected $T_\mathrm{2, f_\text{Q2}=1~GHz}^H$ as extracted from a power law fit to the opaque data markers in~\textbf{f} as a function of the gate voltage sensitivity $(\partial f_\text{Q2}/\partial V_\text{P2})/f_\text{Q2}$ from~\textbf{e}. Coloured markers correspond to the different magnetic field orientations as indicated in~\textbf{e}. Solid black line is a fit of $T_2^H=ax^{\beta}$ to the data, yielding an exponent of $\beta=-1.04(8)$.
    }
\label{fig:esens2}
\end{figure*}

\section*{Charge noise}
The connection between confinement potential of the hole and LH-HH mixing gives rise to a sensitivity of the $g$-tensor to local electric fields~\cite{winkler_spin_2008, wang_optimal_2021}. An electric field modulation will thus induce a variation of the $g$-tensor $\delta\tensor{g}$, which leads to a modulation of the Larmor vector $h\delta\mathbf{f}_\text{Q}=\mu_\text{B}\delta\tensor{g}\cdot\mathbf{B}$. These modulations can be separated into changes parallel (longitudinal) or perpendicular (transverse) to the qubit quantisation axis. The former will change the qubit energy splitting and provide a channel for dephasing due to e.g. charge noise~\cite{piot_single_2022}, while the latter enables driving the qubit through $g$-tensor magnetic resonance ($g$-TMR)~\cite{kato_gigahertz_2003, crippa_electrical_2018, piot_single_2022}. The dependency of $\tensor{g}$ on electric field fluctuations will depend on the direction of the electric field, which we study by considering potential modulations on differently oriented gates.

We first focus on the longitudinal electric field sensitivity $\partial f_\text{Q2}/\partial V_\text{P2}$ of Q2 with respect to its plunger gate, as we expect charge noise to mostly originate from the interfaces and oxides directly above the qubit. We determine the change in qubit frequency from an acquired phase when applying a small voltage pulse $\delta V_i$ to different gate electrodes $i$ during a Hahn echo measurement~\cite{crippa_electrical_2018, piot_single_2022} (see Fig.~\ref{fig:esens2}a-d and Methods). Fig.~\ref{fig:esens2}e shows $\partial f_\text{Q2}/\partial V_\text{P2}$ for different magnetic field orientations, for $f_\text{Q2}=1.36(7)$~GHz. We observe the qubit energy splitting to be most sensitive to electric field fluctuations when $\mathbf{B}$ is in the plane $x'y'$ spanned by the $g$-tensor minor principal axes (indicated by the red dashed line) with $\partial f_\text{Q2}/\partial V_\text{P2}>2$~GHz/V.

If qubit decoherence is limited by fluctuations of the $g$-tensor caused by charge noise, we expect the qubit frequency fluctuations $\delta f_\text{Q}$ to linearly increase with $B$ and to strongly depend on the orientation of $\mathbf{B}$ as governed by the corresponding longitudinal electric field sensitivity. To this end, we perform a Hahn echo experiment and extract the echo coherence times $T_2^\text{H}$ by fitting the data to an envelope exponential decay disregarding nuclear spin effects (see Methods). We plot $T_2^\text{H}$ as a function of the qubit frequency (obtained by varying $B$) in Fig.~\ref{fig:esens2}f for different orientations of $\mathbf{B}$ indicated by the coloured markers in Fig.~\ref{fig:esens2}e. For large enough $B$, we observe a power law dependence of $T_2^\text{H}\propto f_\text{Q}^{-1}$, consistent with a $1/f$ noise spectrum acting on the qubit~\cite{piot_single_2022, cywinski_how_2008} (see Methods). We note that for small $B$, the finite spread of the precession frequencies of the nuclear spin ensemble limits qubit coherence, resulting in a sharp decrease~\cite{bluhm_dephasing_2011} of the extracted $T_2^\text{H}$. Next, we correlate the observed charge-noise limited echo coherence time to the electric field sensitivity of $\tensor g$ for different orientations $\mathbf{B}$. The extracted charge-noise limited $T_2^\text{H}$ obtained at different $(\theta_\mathbf{B}$, $\phi_\mathbf{B})$ and extrapolated to $f_\text{Q2}=1$~GHz is plotted as a function of the measured relative voltage sensitivity $(\partial f_\text{Q2}/\partial V_\text{P2})/f_\text{Q2}$ (Fig.~\ref{fig:esens2}g). The good fit to a power law with exponent $-1$ confirms the dominant source of charge noise originates from the oxide interfaces directly above the qubit (see Supp. Fig.~\ref{fig:gate_comparison} for the noise sensitivity of other gates).

\begin{figure*}[htp]
\includegraphics{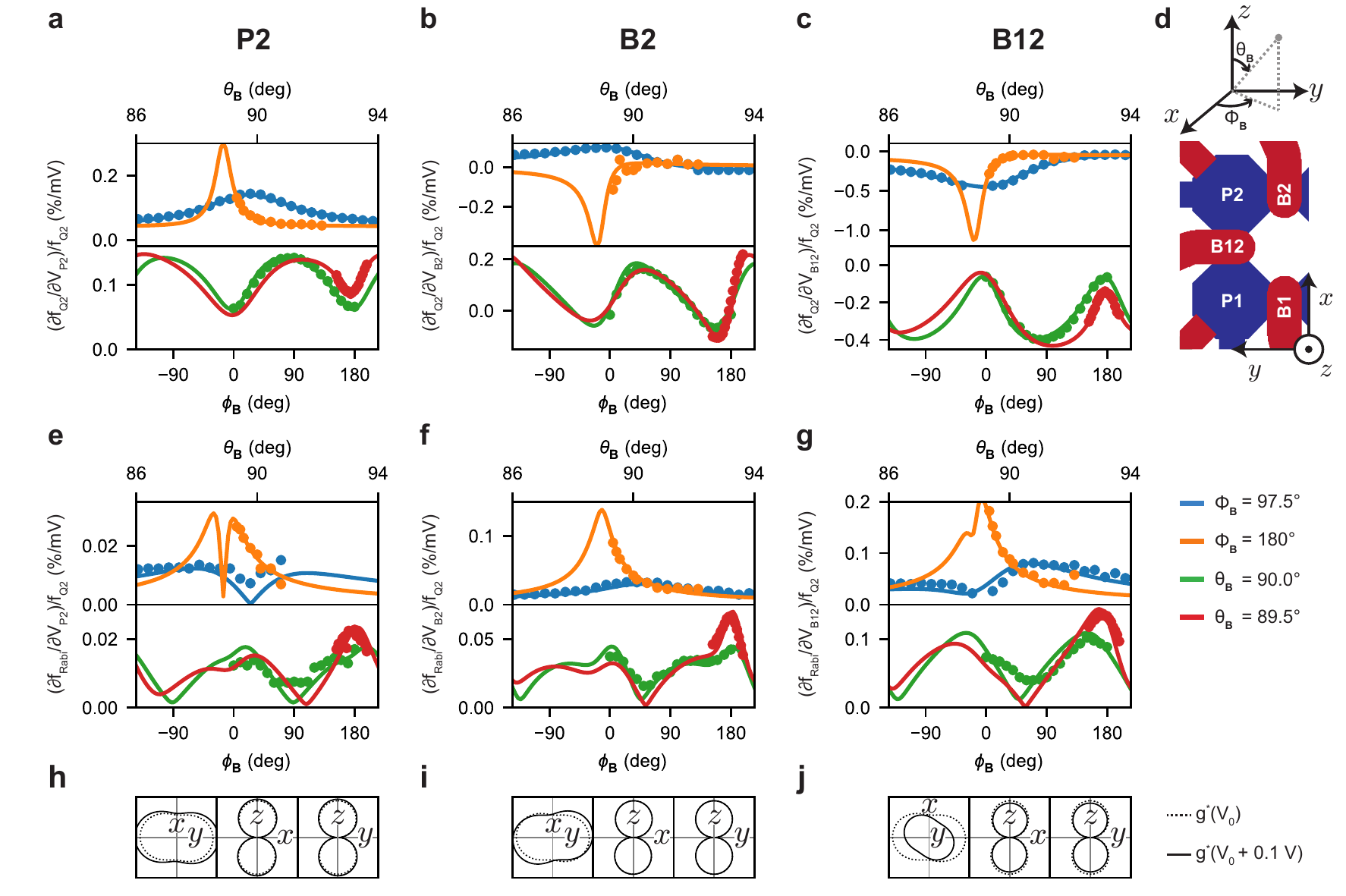}
\caption{\textbf{Reconstruction of $\bm{\partial\tensor{g}/\partial V_i}$ for differently oriented electrostatic gates.}
    \textbf{a}-\textbf{c}, Relative voltage sensitivity of the energy splitting $(\partial f_\text{Q2}/\partial V_i)/f_\text{Q2}$ of Q2 for a voltage excitation on gates P2 (\textbf{a}), B2 (\textbf{b}), and B12 (\textbf{c}). Top panels correspond to sweeps of the magnetic field elevation $\theta_\mathbf{B}$, while bottom panels correspond to sweeps of the in-plane angle $\phi_\mathbf{B}$. The solid lines correspond to projections of the  $\partial\tensor{g}/\partial V_i$ fitted to the data.
    \textbf{d}, Schematic illustration of the qubit layout indicating the different electrostatic gates.
    \textbf{e}-\textbf{g}, Relative Rabi frequency of $(\partial f_\text{Rabi}/\partial V_i)/f_\text{Q2}$ of Q2 for a drive voltage excitation $V_i$ on gates P2 (\textbf{e}), B2 (\textbf{f}), and B12 (\textbf{g}). Solid lines correspond to the projection of the $\partial\tensor{g}/\partial V_i$ as fitted to the data in panels~\textbf{a}-\textbf{c}.
    \textbf{h}-\textbf{j}, Cross-section of the change of the $g$-tensor in the $xy$, $xz$, and $yz$-planes of the magnet frame. Dashed lines correspond to cross-sections of $\tensor{g}$, while solid lines represent $\tensor{g}+\delta\tensor{g}_i(0.1 V)$, for gates P2 (\textbf{h}), B2 (\textbf{i}), and B12 (\textbf{j}).
    }
\label{fig:esens_fitting}
\end{figure*}

To get a full understanding of the mechanism underlying the electric modulation of the $g$-tensor, we reconstruct $\partial\tensor{g}/\partial V_i$ for Q2 as a function of the voltage applied to plunger gate P2, and two neighbouring barrier gates B2 and B12, oriented at a $90\degree$ angle to each other (Fig.~\ref{fig:esens_fitting}d). We measure $(\partial f_\text{Q2}/\partial V_i)/f_\text{Q2}$ for selected magnetic field orientations, that together with the previously extracted $\tensor{g}$ allows fitting $\partial\tensor{g}/\partial V_i$ (see Methods). All measurements are performed at constant $f_\text{Q2}=225$~MHz and we show the measured relative electric potential sensitivity and corresponding fits in Fig.~\ref{fig:esens_fitting}a-c. The extracted fit parameters are detailed in Supp.~Table~\ref{tab:dg_tensor}. To illustrate what happens to the $g$-tensor as the gates are pulsed, we sketch the cross-sections of $\tensor{g}$ and $\tensor{g}+\delta\tensor{g}_i(\delta V_i)$ in the $xz$, $yz$, and $xy$ planes of the magnet frame for $\delta V_i=100$~mV in Fig.~\ref{fig:esens_fitting}h-j. We observe that the plunger gate directly above the qubit mostly scales the $g$-tensor principle axes (`breathing'), while the neighbouring barrier gates also induce a rotation of the $g$-tensor (Supp. Table~\ref{tab:dg_tensor}). A true sweet spot to noise originating near gate $i$ exists when $\partial f_\text{Q}/\partial V_i=0$. We only find such a zero crossing for potentials applied to side gate B2, as visible in Fig.~\ref{fig:esens_fitting}b (see Supp. Fig.~\ref{fig:fitprojections} for the full $\theta_\mathbf{B}\phi_\mathbf{B}$-projections). For voltage fluctuations applied to gates P2 and B12, we find that an improvement in the electric field sensitivity is possible, but no true sweet spot exists for any $(\theta_\mathbf{B}$, $\phi_\mathbf{B})$. These effects are dominated by the dynamic tilting of $\tensor{g}$, which we believe to be caused by hole wave function moving in a local strain gradient~\cite{corley-wiciak_nanoscale_2023, liles_electrical_2020, abadillo-uriel_hole_2022}, not taken into account by previous models~\cite{wang_modelling_2022}.

While the longitudinal component of the $g$-tensor modulation leads to decoherence, the transverse part enables an electric drive of the qubit through $g$-TMR. Therefore, our reconstruction of $\partial\tensor{g}/\partial V_i$ allows us to compare the expected Rabi frequency from $g$-TMR with the observed Rabi frequency. We measure the angular dependence of the Rabi frequency of the qubit, for a resonant electric drive with amplitude $V_i$ applied to either gate P2, B2, or B12 and extract $(\partial f_\text{Rabi}/\partial V_i)/f_\text{Q2}$. The results, shown in Fig.~\ref{fig:esens_fitting}e-g, reveal a striking agreement between the measured Rabi frequency and the expected drive due to the $g$-TMR (see Methods for details). The agreement between the data and the projection of $\partial\tensor{g}/\partial V_i$, both in absolute size and magnetic field dependence, confirms that the main driving mechanism of planar germanium hole qubits is in fact $g$-TMR.

\begin{figure*}[htp]
\includegraphics{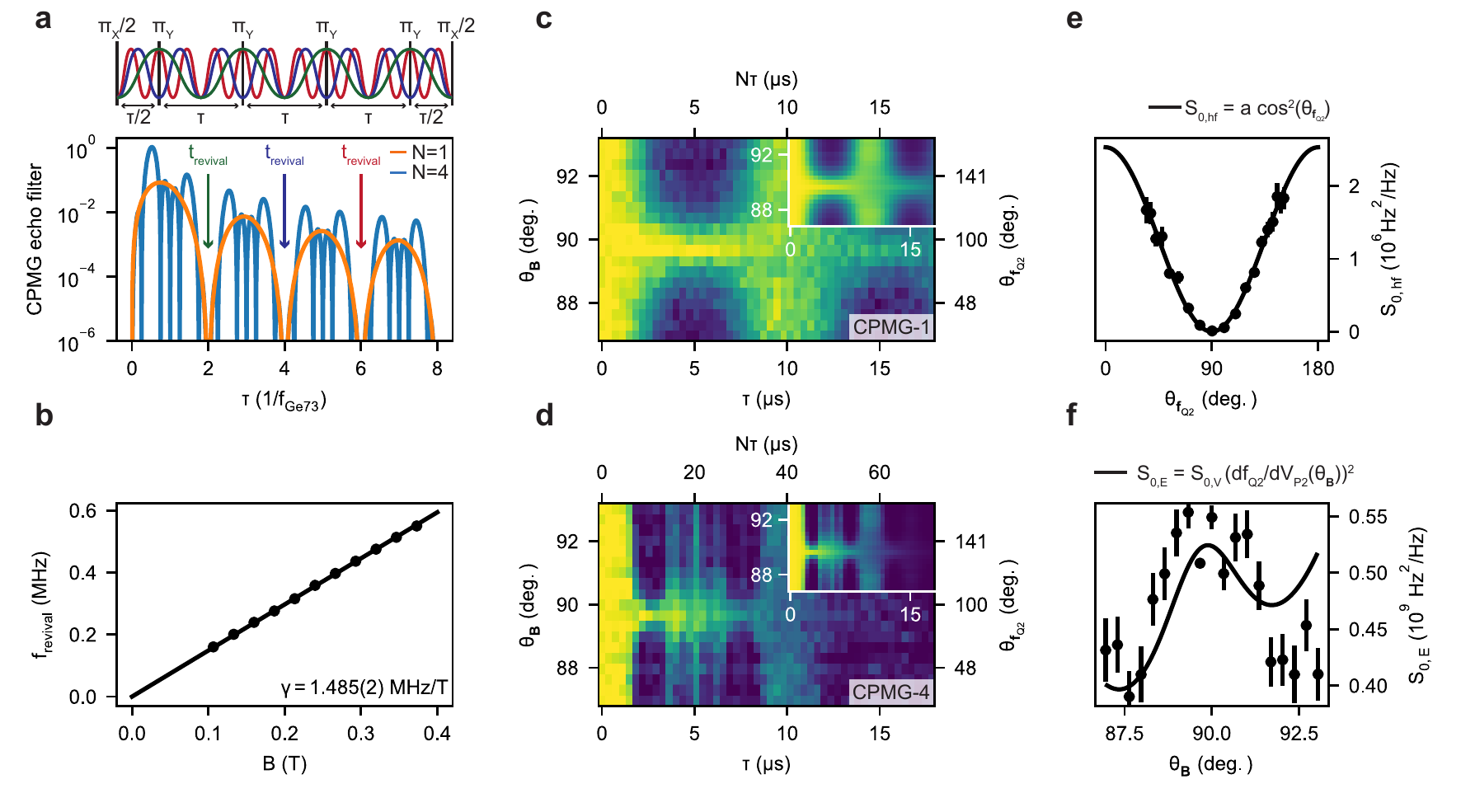}
\caption{\textbf{Collapse-and-revival of qubit coherence due to hyperfine interaction.}
    \textbf{a}, Filter function of the CPMG pulse sequence, for $N=1$ and 4 decoupling pulses, illustrating full suppression of noise with a characteristic frequency $f=n/(2\tau)$, with $n$ any integer.
    \textbf{b}, Extracted revival frequency as a function of the magnetic field strength $B$, full data shown in Supp. Fig.~\ref{fig:cpmg_field}. We extract a gyromagnetic ratio of the $^{73}$Ge nuclear spin of $\gamma=1.485(2)$~MHz/T.
    \textbf{c},~\textbf{d}, Normalized charge sensor signal for a CPMG sequence with respectively 1 (\textbf{c}), and 4 (\textbf{d}) decoupling pulses, as a function of the spacing between two subsequent decoupling pulses $\tau$ and $\theta_\mathbf{B}$. $N\tau$ is the total evolution time. $\phi_\mathbf{B}=97.5\degree$ and $B=133$~mT. The inset displays the fit to the data from which we extract $S_{0,\text{hf}}(\theta_\mathbf{B})$ and $S_{0,\text{E}}(\theta_\mathbf{B})$.
    \textbf{e}, The extracted strength of the hyperfine interaction as a function of $\theta_\mathbf{f_\text{Q}}$. The black line is a fit of the data to $a\cos^2(\theta_\mathbf{f_\text{Q}})$, with $a=2.5\cdot10^{6}~\text{Hz}^2/\text{Hz}$. Error bars indicate $1\sigma$ of the fit.
    \textbf{f}, The extracted strength of the $1/f$ noise at $1$~Hz. The black line is a fit of the data to $S_{0,\text{E}}=S_{0,V}\cdot(\partial f_\text{Q2}/\partial V_\text{P2}(\theta_\mathbf{B}))^2$, with $\partial f_\text{Q2}/\partial V_\text{P2}(\theta_\mathbf{B})$ the electric field sensitivity of the qubit frequency to the top gate voltage as extracted from Fig.~\ref{fig:esens_fitting} and $S_{0,V}=6.1\cdot10^{-10}~V^2/\text{Hz}$ the only fit parameter. Error bars indicate $1\sigma$.
    }
\label{fig:nuclear_combined}
\end{figure*}

\section*{Hyperfine interaction}
\label{sec:hyperfine}
Our qubits are defined in a natural germanium quantum well, where with  a concentration of 7.7~\%, $^{73}$Ge is the only isotope with non-zero nuclear spin. As a result, the hole wave function overlaps with ${\sim}10^6$ nuclear spins (see Methods), leading to a fluctuating Overhauser field acting on the hole spin. One can separate the contributions of the Overhauser field into longitudinal and transverse components with respect to the quantisation axis of the nuclear spins~\cite{bluhm_dephasing_2011}. While temporal fluctuations of both components can lead to qubit dephasing, longitudinal field fluctuations are mainly caused by the quasi-static dipole-dipole interaction between nuclear spins~\cite{cywinski_electron_2009, chekhovich_nuclear_2013} and can easily be echoed out. However, the transverse part contains a spectral component at the Larmor frequency of the nuclear spins, that leads to a collapse-and-revival of coherence when performing spin echo experiments, as predicted in Refs.~\cite{cywinski_electron_2009, cywinski_pure_2009} and observed in GaAs~\cite{bluhm_dephasing_2011} and germanium~\cite{lawrie_spin_2022}.

The hyperfine interaction between heavy hole states and nuclear spins is expected to be highly anisotropic~\cite{fischer_spin_2008}, unlike the isotropic contact hyperfine interaction observed for conduction band electrons. In fact, for the $^{73}$Ge isotope, the Ising term (out-of-plane, $\propto s_zI_z$) is numerically estimated to be ${\sim}50$ times larger than in-plane ($\propto s_xI_x, s_yI_y$) components~\cite{philippopoulos_hyperfine_2020}. As a result, hyperfine interaction between the heavy hole and the surrounding nuclear spin bath is expected to be negligible for an in-plane magnetic field~\cite{fischer_spin_2008, philippopoulos_hyperfine_2020}. To study the hyperfine anisotropy for planar germanium qubits, we perform a Carr-Purcell-Meiboom-Gill (CPMG) experiment, which constitutes an effective band pass filter for the noise acting on the qubit with a frequency $f=1/\tau$ (Fig.~\ref{fig:nuclear_combined}a). We apply CPMG sequences with $N=1$, $2$, $4$, and $8$ decoupling pulses to Q2 and measure the spin state as a function of the free evolution time $\tau$ between the $\text{Y}_\pi$-pulses, as shown in Fig.~\ref{fig:nuclear_combined}c,d for $N=1$ and $N=4$ (data for $N=2$, $N=8$ in Supp. Fig.~\ref{fig:fig_nuclear_rest}c,d). We observe the expected collapse-and-revival of the coherence and find $f_\text{revival}=\gamma|\mathbf{B}|$ with $\gamma=1.485(2)$~MHz/T (Fig.~\ref{fig:nuclear_combined}b), in good agreement with the gyromagnetic ratio of the $^{73}$Ge nuclear spin $\gamma_\text{Ge-73}=1.48$~MHz/T, confirming its origin. 

Following Refs.~\cite{uhrig_keeping_2007, cywinski_how_2008}, we fit the data using the formalism developed to describe decoherence of dynamically decoupled qubits suffering from dephasing noise with a given noise spectrum. We assume a noise spectrum $S_{f_q}$ acting on the qubit, consisting of a $1/f$ part caused by charge noise, as well as a sharp spectral component at the precession frequency of the $^{73}$Ge nuclear spins (see Methods). We extract the strength of the nuclear noise $S_{0,\text{hf}}$ and plot this as a function of the elevation of the Larmor vector $\theta_{\mathbf{f}_\text{Q2}}=\arccos\left(\frac{\mathbf{f_\text{Q2}}\cdot\hat{z}}{f_\text{Q2}}\right)$ (Fig.~\ref{fig:nuclear_combined}e). We find that the data closely follows a relation $S_{0,\text{hf}}\propto\cos^2\left(\theta_{\mathbf{f}_\text{Q2}}\right)$, providing strong experimental evidence of the predicted Ising coupling~\cite{fischer_spin_2008, prechtel_decoupling_2016, philippopoulos_hyperfine_2020}. As a result, there exists a sweet plane approximately spanned by the $x'y'$ axes of the $g$-tensor, where the qubit is mostly insensitive to nuclear spin noise. The finite width of the hyperfine distribution of $\sigma_\text{Ge-73}=9-16$~kHz, results in a loss of qubit coherence at small $B$, as seen in Fig.~\ref{fig:esens2}f. This line width is several orders of magnitude larger than expected for a single $^{73}$Ge spin~\cite{kaufmann_73ge_1971}, but is in good agreement with values previously observed in Ge~\cite{lawrie_spin_2022} and could be caused either by interactions between the nuclei or by the quadrupolar splitting present in the $^{73}$Ge isotope.

Finally, assuming all charge noise to originate near P2, we fit the extracted $S_{0,E}$ (Fig.~\ref{fig:nuclear_combined}f) to $\partial f_\text{Q2}/\partial V_\text{P2}$ as measured in Fig.~\ref{fig:esens_fitting} and find an effective electric noise power spectral density of $S_{V}=610~\mu V^2/\mathrm{Hz}$ at 1~Hz, corresponding to an effective voltage noise of $25~\mu V/\sqrt{\mathrm{Hz}}$ on P2. Using the estimated plunger gate lever arm $\alpha_\mathrm{P}=7.4\%$ (Supp. Fig.~\ref{fig:leverarm}) we extract a charge noise level of $1.9~\mu eV/\sqrt{\mathrm{Hz}}$, in good agreement with charge noise measurements on similar devices~\cite{lodari_low_2021}.

\begin{figure*}[htp]
\includegraphics{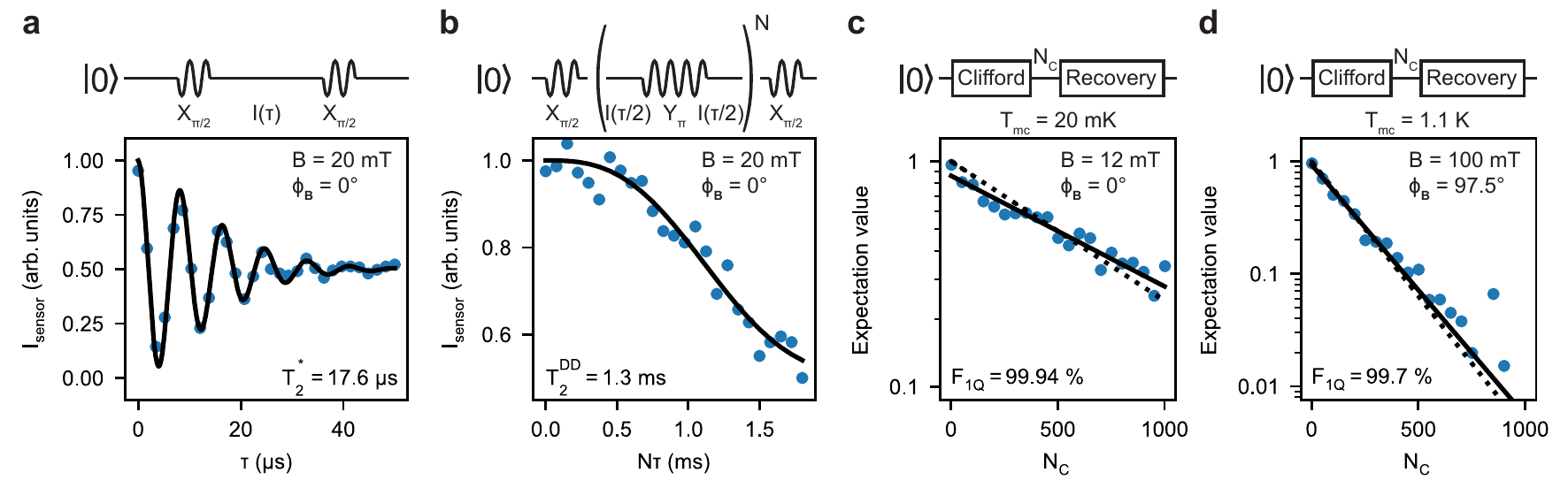}
\caption{\textbf{Coherence figures at low magnetic field in the hyperfine sweet spot.}
    \textbf{a}, Free induction decay coherence as measured through a Ramsey experiment. The data constitute of an average of 10 traces, for a total integration time of approximately 5 minutes (12-hour dataset Supp.~Fig.~\ref{fig:ramsey_20mT}) and we find a coherence time of $T_2^*=17.6~\mu$s.
    \textbf{b}, CPMG dynamical decoupling coherence, as measured for a sequence with 250 refocusing pulses. We find a coherence time of $T_2^\text{DD}=1.3$~ms.
    \textbf{c}, Randomized benchmarking of the performance of Q2. The solid line is a fit of the data to $P=a\exp(-(1-2F_\text{c})N_\text{C})$, from which we extract a single qubit gate fidelity of $F_\text{g}=99.94$~\%. The reduced visibility for larger $N_\text{C}$ is caused by the readout being affected by the large number of pulses applied to the gate, but does not affect the extracted fidelity (see Methods), as indicated by the dashed line where we fix $a=1$.
    \textbf{d}, Randomized benchmarking at a fridge temperature of $T=1.1$~K for Q2. We now operate in the joint Q1-Q2 hyperfine sweet spot at $\phi_\mathbf{B}=97.5~\degree$. We extract a single qubit gate fidelity of $F_\text{g}=99.7$~\%.
    }
\label{fig:coherence}
\end{figure*}

\section*{Sweet-spot operation}
The detailed understanding of the hole qubit coherence for different magnetic field orientations, allows to select an optimal operation regime. For any magnetic field orientation away from the hyperfine sweet plane, nuclear spin noise limits qubit coherence in natural germanium samples. However, the slight but significant tilt between the two qubit $g$-tensors limits this further to a single spot where the two circles intersect: $\phi_\mathbf{B}=97.5~\degree$ and $\theta_\mathbf{B}=89.7~\degree$ for this device (see Supp. Fig.~\ref{fig:fig_nuclear_rest}). The existence of such common hyperfine sweet spots is not guaranteed for larger qubit systems when the individual qubit $g$-tensors slightly differ. Furthermore, we observe that this hyperfine sweet plane coincides with the hot spots for charge-induced decoherence (Supp. Fig.~\ref{fig:fitprojections}), preventing full employment of charge noise sweet spots. In fact, we estimate charge-noise limited coherence times and quality factors to be improved by about an order of magnitude for the optimal magnetic field orientation. This showcases the need for isotopically purified materials, despite the Ising type hyperfine interaction of the heavy hole.

For our device, we aim to optimize the coherence of Q2 by lowering the magnetic field strength and operate along the hyperfine sweet plane of Q2, with $\phi_\mathbf{B}=0\degree$ to strike a balance between low charge noise sensitivity and high operation speed. We first assess the free induction decay coherence time by performing a Ramsey experiment (Fig.~\ref{fig:coherence}a). We set $B=20$~mT, such that $f_\text{Q2}\approx21$~MHz and $f_\text{Rabi}=1$~MHz and find $T_2^*=17.6~\mu$s, which is about an order of magnitude larger than shown previously for germanium hole qubits~\cite{lawrie_spin_2022}. We can further extend the coherence by using dynamical decoupling and find coherence times beyond 1~ms (Fig.~\ref{fig:coherence}b). Operation at low magnetic field also has implications on the speed of single qubit operations, as these are expected and observed to scale with $B$ (Eq.~\ref{eq:dfrdV} in Methods). Single qubit gate performance is ultimately governed by the ratio of the operation time and the coherence time, and should thus in principle be preserved even at low magnetic fields. To test this, we perform randomized benchmarking, with a Clifford group based on $X_\pi$ and $X_{\pi/2}$ pulses and virtual $Z$ updates (see Supp. Table~\ref{tab:RB_cliff}). We find an optimal average single qubit gate fidelity (with 0.875 physical gates per Clifford) of $99.94$~\% at $B=12$~mT (Fig.~\ref{fig:coherence}c), well above $99$~\%. Furthermore, we find that the fidelity remains significantly higher than $99$~\% when operating our qubits at an elevated temperature of $T=1.1$~K, where more cooling power is available (Fig.~\ref{fig:coherence}d). Lowering the qubit frequency thus opens a path to increase qubit coherence, while maintaining high single-qubit gate performance. This can provide a potential avenue to improve two-qubit gate performance, which has typically been limited by the comparatively short coherence time of the germanium hole qubit~\cite{hendrickx_four-qubit_2021, van_riggelen_phase_2022}.

\section*{Conclusions}
In summary, we report on a fully electrically controlled two-qubit system defined by single hole spins in a strained germanium quantum well. The hole $g$-tensor of both qubits is characterized, revealing a strong anisotropy with respect to the heterostructure growth direction. The two qubit $g$-tensors are remarkably similar and vary by less than $10$~\%, indicative of a high degree of uniformity of the electrostatic confinement. However, the small tilt ($\delta\theta\approx1~\degree$) combined with the large anisotropy of $\tensor{g}$ leads to measurable effects, in particular for magnetic field orientations in proximity to the $g$-tensor minor principle axes. The slight tilt of the $g$-tensor is likely the result of local strain gradients and could thus be controlled through material and gate stack optimization, or by modifying the LH-HH mixing, defined by material stoichiometry~\cite{lodari_lightly_2022} and quantum dot confinement~\cite{bosco_squeezed_2021}.

The $g$-tensor anisotropy is also reflected in the qubit sensitivity to electric field fluctuations. We find that $\tensor{g}$ breathes and tilts under electric field fluctuations, leading to charge-noise induced decoherence, but also enabling qubit control through $g$-TMR, both strongly anisotropic in strength with respect to the magnetic field orientation. Furthermore, also the hyperfine interaction between the qubit and the $^{73}$Ge nuclear spin bath is extremely anisotropic and only suppressed when the qubit quantisation axis aligns with the quantum well plane. As a result, the hyperfine interaction is detrimental to qubit coherence for any $\mathbf{B}\nparallel x'y'$. When the nuclear spin noise can be mitigated, we find qubit coherence to be limited by charge noise with a $1/f$ power spectrum, such that coherence times are inversely correlated to the qubit energy splitting and its electric field sensitivity. The hyperfine interaction hinders leveraging of the electric field sensitivity sweet spots that would enable a significant further improvement to qubit coherence, underpinning the need for isotopic purification of the germanium quantum well~\cite{itoh_high_1993}. Finally, we find that qubit coherence can be substantially increased by operating in the low-field regime, while maintaining high-fidelity single qubit control with a gate fidelity well above the fault tolerant threshold, even at operation temperatures above 1~K. This understanding of the dominant decoherence mechanisms and sweet spots for hole spins is key for the future design and operation of large-scale, high-fidelity spin qubit arrays.


\section*{Methods}
\setcurrentname{Methods}
\label{sec:methods}
\subsection*{Device fabrication}
The quantum dot device is fabricated on a Ge/SiGe heterostructure consisting of a 20-nm-thick quantum well buried 48~nm below the wafer surface, grown in an industrial reduced-pressure chemical vapour deposition reactor~\cite{bedell_low-temperature_2020}. The virtual substrate consists of a strain-relaxed germanium layer on a silicon wafer and multiple layers with increasing silicon content to reach the Si$_{0.2}$Ge$_{0.8}$ stoichiometry used for the quantum well barriers. Ohmic contacts to the quantum well are defined by in-diffusion of Pt at a temperature of $300\degree$~C. We note that in the device used for this work, the Pt-silicide did not diffuse in deep enough to reach the quantum well, resulting in a larger contact resistance ($\sim$M$\Omega$). Electrostatic gates are defined using electron beam lithography and lift-off of Ti/Pd (20~nm), separated by thin (7~nm) layers of atomic layer deposited SiO$_2$.

\subsection*{Experimental setup}
All measurements are performed in a Bluefors LD400 dilution refrigerator with a base temperature of $T_\text{mc}=10$~mK. The sample is mounted on a QDevil QBoard circuit board, and static biases are applied to the gates using a QDevil QDAC through dc looms filtered using a QDevil QFilter at the millikelvin stage of our fridge. In addition, all plunger and barrier gates are also connected to coaxial lines through on-PCB bias-tees. All rf lines are attenuated by 10~dB at the 4K stage and an additional 3~dB at the still. We use Tektronix AWG5204 arbitrary waveform generators (AWGs) to deliver fast voltage excitation pulses to the quantum dot gates. Furthermore, we use the AWGs to drive the vector  input of a Rohde~\&~Schwarz SGS100A source to generate microwave control signals when $f_\text{Q}>500$~MHz. For experiments when $f_\text{Q}<500$~MHz, we directly synthesize the qubit drive pulses using the AWG. Unfortunately, the coaxial line connected to gate P1 was defective at the time of the experiments. To enable fast pulsing throughout the charge stability diagram of the double quantum dot, we applied pulses to the coaxial line connected to RB1 the reservoir side gate of Q1 (see Fig.~\ref{fig:dots}) instead and account for the difference in dot-gate capacitance between P1 and RB1. The independent control over the dc voltage on RB1 and P1 still allows to select a reservoir tunnel rate suitable for the experiments.

The qubits are read out using a charge sensor defined in the lower channel of the four quantum dot device. We tune the device to form a single quantum dot underneath the central gate SP, with the tunnel rates being controlled by SB1 and SB2 as defined in Fig.~\ref{fig:dots} of the main text. We measure the sensor conductance using a pair of Basel Precision Instruments (BasPI) SP983c IV-converters with a gain of $10^6$ and a low-pass output filter with a cut-off frequency of 30~kHz and applying a source-drain bias excitation of $V_\text{SD}=300-800~\mu$V. We directly extract the differential current using a BasPI SP1004 differential amplifier with a gain of $10^3$ and record the signal using an Alazar ATS9440 digitizer card.

An external magnetic field is applied through an American Magnetics three-axis magnet with a maximum field of 1/1/6 Tesla in the $xyz$ direction and a high-stability option on all coils. We note that due to an offset $z=2.78$~cm of the sample with respect to the $xy$ coil centres, a correction of $-11.2$\% is applied to $B_x$ and $B_y$ as following from a simulation of the magnet coil fields. As the sample is correctly centred with respect to the $z$ solenoid, no off-diagonal components of the applied magnetic field are present (i.e. $B_{x-\text{coil}}\parallel x$, $B_{y-\text{coil}}\parallel y$, and $B_{z-\text{coil}}\parallel z$). The correctly observed gyromagnetic ratio of the $^{73}$Ge nuclear spin confirms the accuracy of this correction. Small common rotations of the Q1 and Q2 $g$-tensor rotations may occur due to imperfect planar mounting of the sample. Finally, we note that our magnet coils typically show a few mT of hysteresis, which becomes significant at very low fields. To ensure operation in a hyperfine sweet spot, we sweep $\theta_\mathbf{B}$ before every measurement in Fig.~\ref{fig:coherence} and locate the sweet plane by minimizing the qubit frequency as a function of $\theta_\mathbf{B}$.

\subsection*{Virtual gate matrices}
To compensate for the cross capacitance between the different electrostatic gates and the quantum dots, we define a set of virtual gates~\cite{hensgens_quantum_2017}:
\begin{equation*}
\begin{pmatrix}
V_\text{P1} \\ V_\text{P2} \\ V_\text{P3} \\ V_\text{P4} \\ V_\text{B12}
\end{pmatrix}= 
\begin{pmatrix} 
1&-0.28&0&0&-1.65\\
-0.18&1&0&0&-1.30\\
0&-0.11&1&0&0.10\\
-0.11&0&0&1&0.10\\
0&0&0&0&1\\
\end{pmatrix}
\begin{pmatrix} 
V_{\overline{\text{P1}}} \\
V_{\overline{\text{P2}}} \\
V_{\overline{\text{SB2}}} \\
V_{\overline{\text{SB1}}} \\
V_{\overline{\text{B12}}} \\
\end{pmatrix}
\end{equation*}
with G$_i$ the real gate voltage, and $\overline{\text{G}_i}$ the virtual gate voltage, which leaves the chemical potential of the other quantum dots unchanged. Furthermore, we define a second pair of axes detuning $\epsilon$ and on-site energy U, as illustrated in Fig.~\ref{fig:dots}b of the main text:
\begin{equation*}
\begin{pmatrix}
V_{\overline{\text{P1}}} \\ V_{\overline{\text{P2}}}
\end{pmatrix}= 
\begin{pmatrix} 
-0.5&0.5\\
0.5&0.5\\
\end{pmatrix}
\begin{pmatrix} 
V_\epsilon \\
V_\text{U} \\
\end{pmatrix}
\end{equation*}

\subsection*{Pauli spin blockade readout}
To overcome rapid spin relaxation as mediated by the spin-orbit interaction~\cite{danon_pauli_2009}, we make use of charge latching, where we tune the tunnel rates between each dot and its respective reservoir to be asymmetric $t_\text{Q2}\ll t_\text{Q1}$. By pulsing across the extended (1,1)-(0,1) charge transition line, we can latch the blocking (1,1) states into a (0,1) charge state~\cite{harvey-collard_high-fidelity_2018, hendrickx_four-qubit_2021}, with a characteristic decay time to the (0,2) ground state governed by $t_\text{Q2}$. Furthermore, the spin-orbit interaction introduces a coupling between the $\ket{T(1,1)}$ and $\ket{S(0,2)}$ states, resulting in the presence of an anticrossing between the $\ket{\downarrow\downarrow}$ and the $\ket{S(0,2)}$ states. As a result, depending on the sweep rate across the interdot transition line, as well as the orientation of the external magnetic field $B$, we observe either parity or single-state readout~\cite{seedhouse_pauli_2021, hendrickx_four-qubit_2021}. We typically operate the device in single-state readout by sweeping fast across the anti-crossing, unless this was prohibited due to the finite bandwidth of our setup with respect to the different tunnel rates.

\subsection*{Fitting procedure of the $\bm{g}$-tensor}
The $g$-tensor of the device can be described as a rotated diagonal matrix:
\begin{equation}
\label{eq:gtensor}
    \tensor{g}=R(\phi,~\theta,~\zeta)\diag{(g_{x'}, g_{y'}, g_{z'})}R^{-1}(\phi,~\theta,~\zeta)
\end{equation}
where Euler angles $\phi$, $\theta$, and $\zeta$ define the successive intrinsic rotations around the $zyz$ axes. The rotation matrix $R$ is thus defined as:
\begin{widetext}
\begin{equation}
\label{eq:rotmat}
    R(\phi,~\theta,~\zeta)=
    \left(
\begin{array}{ccc}
\cos (\phi ) \cos (\theta ) \cos (\zeta )-\sin (\phi ) \sin (\zeta ) & -\sin (\phi ) \cos (\zeta )-\cos (\phi )
   \cos (\theta ) \sin (\zeta ) & \cos (\phi ) \sin (\theta ) \\
 \sin (\phi ) \cos (\theta ) \cos (\zeta )+\cos (\phi ) \sin (\zeta ) & \cos (\phi ) \cos (\zeta )-\sin (\phi )
   \cos (\theta ) \sin (\zeta ) & \sin (\phi ) \sin (\theta ) \\
 -\sin (\theta ) \cos (\zeta ) & \sin (\theta ) \sin (\zeta ) & \cos (\theta ) \\
\end{array}
\right)
\end{equation}
\end{widetext}

The $g$-tensor can thus be reconstructed by measuring the qubit energy splitting $hf_\text{Q}$ for different orientations of the magnetic field $\mathbf{B}$. We measure $f_\text{Q}$ for various magnetic field orientations $(\theta_\mathbf{B}, \phi_\mathbf{B})$ and fit the data to:
\begin{equation}
    \label{eq:zeeman}
    hf_\text{Q} = \abs{\mu_B\tensor{g}\cdot\mathbf{B}}
\end{equation}
using $\tensor{g}$ as defined in Eqs.~\ref{eq:gtensor}-\ref{eq:rotmat} and $g_{x'}$, $g_{y'}$, $g_{z'}$, $\phi$, $\theta$, and $\zeta$ as fitting parameters. The data used for the fitting include but are not limited to the data presented in Fig.~\ref{fig:g_tensor} of the main text. All magnetic field orientations at which $f_\text{Q}$ is measured are shown in Supp.~Fig.~\ref{fig:edsr_spectra}c. These field orientations $(\theta_\mathbf{B}, \phi_\mathbf{B})$ are selected to enable a reliable fit of the $g$-tensor, with the error on the different parameters indicated in Fig.~\ref{fig:g_tensor}h of the main text.

\subsection*{Fitting procedure of the charge-noise limited coherence}
We measure the qubit coherence by extracting the Hahn echo coherence time, which is insensitive to quasi-static noise and experimental parameters such as the integration time. We measure the normalized charge sensor current as a function of the total free evolution time $2\tau$ and observe two different regimes (see Supp. Fig.~\ref{fig:hahn_example}). In the first (Supp. Fig.~\ref{fig:hahn_example}a), the echo decay follows an exponential decay and we fit the data to $I_\text{SD}=\exp(-(2\tau/T_2^\text{H})^\alpha)$, with the exponent $\alpha$ left free as a fitting parameter (Supp. Fig.~\ref{fig:hahn_example}a). However, for magnetic field orientations where the echo decay is dominated by the nuclear spin induced decoherence ($\mathbf{B}\nparallel x'y'$), we extract the envelope coherence $T_2^\text{H}$ by fitting the envelope of the nuclear spin induced collapse-and-revival (Supp. Fig.~\ref{fig:hahn_example}a)~\cite{bluhm_dephasing_2011} to $I_\text{SD}=\exp(-(2\tau/T_2^\text{H})^\alpha)/|1-a_0\cos(2\pi f_\text{Ge-73}\tau)|^2$, with $a_0$ and $\alpha$ free fitting parameters and $f_\text{Ge-73}=\gamma_\text{Ge-73}B$, as discussed further in the main text.

The exponent of the dependence of the Hahn echo coherence time on both $\partial f_\text{Q}/\partial V_i/f_\text{Q}$ and $f_\text{Q}$ (Fig.~\ref{fig:esens2}f,g of the main text), is related to the colour of the electric noise spectrum. Assuming charge noise with a power law noise spectrum $S\propto f^{\alpha}$ acting on a qubit and following the filter formalism from Refs.~\cite{piot_single_2022, cywinski_how_2008}, we find:
\begin{equation}
    T_2^H\propto\left(\frac{\frac{\partial f_\text{Q}}{\partial V_i}}{f_\text{Q}}(\theta_\mathbf{B},\phi_\mathbf{B})\cdot f_\text{Q}(B)\right)^{\frac{2}{\alpha-1}}
\end{equation}
Therefore, both the dependence of $T_2^H$ on the qubit frequency (by varying $B$, Fig.~\ref{fig:esens2}f) and on the electric field sensitivity (by varying $\theta_\mathbf{B}$ and $\phi_\mathbf{B}$, Fig.~\ref{fig:esens2}g) should obey a power law with exponent $\beta=\frac{2}{\alpha-1}$. From this we can derive the noise exponent $\alpha=\frac{2}{\beta}+1$, such that $\alpha=-1$ if $\beta=-1$.

To obtain the expected charge-noise limited $T_2^H$ at $f_\text{Q2}=1$~GHz, we fit a power law $T_2^H=T_{2}^H[1~\text{GHz}]\cdot 1~\text{GHz}/f_\text{Q2}$ to the data in Fig.~\ref{fig:esens2}f where $B>B_\text{hyperfine}$ (opaque markers). Here, $B_\text{hyperfine}$ indicates the magnetic field strength below which the finite spread of the nuclear spin precession frequencies limits qubit coherence \cite{bluhm_dephasing_2011}. 

Because of the limited maximum field strength we can apply along the $x$ and $y$ axis $B_{\text{max},x}=B_{\text{max},y}=1$~T, the electric field sensitivity for the pink data point is obtained at a lower qubit frequency $f_\text{Q2}=785$~MHz and extrapolated to $f_\text{Q2}=1.36$~GHz.

\subsection*{Fitting procedure of the hyperfine noise}
We follow the method presented in Refs.~\cite{uhrig_keeping_2007, uhrig_exact_2008, cywinski_how_2008} and assume a noise spectrum acting on the qubit consisting of a $1/f$ noise spectrum caused by a large number of charge fluctuators and a Gaussian line caused by the hyperfine interaction with the precession of the $^{73}$Ge nuclear spins:
\begin{equation}
\begin{split}
    S_{f_q}(f, \mathbf{B}) &= S_\text{hf}(f, \mathbf{B}) + S_{E}(f, \mathbf{B}) \\
    S_\text{hf}(f, \mathbf{B}) &= S_{0, \text{hf}}(\mathbf{B})\exp\left(-\frac{f-\gamma_\text{Ge-73}B}{2\sigma_\text{Ge-73}^2}\right) \\
    S_E(f, \mathbf{B}) &= \frac{S_{0,E}}{f}(\mathbf{B}) = \frac{S_{0, V}\left(\frac{\partial f_\text{Q}}{\partial V_\text{P2}}(\mathbf{B})\right)^2}{f}
\end{split}
\end{equation}
Here, $S_{0,\text{hf}}(\mathbf{B})$ defines the effective strength of the nuclear spin noise acting on the qubit, which can be related to the hyperfine coupling constants as detailed below. Furthermore, $\gamma_\text{Ge-73}=1.48$~MHz/T is the $^{73}$Ge gyromagnetic ratio and $\sigma_\text{Ge-73}$ represents the finite spread of the $^{73}$Ge precession frequencies. The charge noise acting on the qubit is most likely originating from charge traps in the interfaces and oxides directly above the qubit, so we model its coupling as coming from the qubit plunger gate, in agreement with what we find in Fig.~\ref{fig:esens2} of the main text. $S_{0, V}$ is the effective voltage noise power spectral density and $\frac{\partial f_\text{Q2}}{\partial V_\text{P2}}(\mathbf{B})$ is the sensitivity of the qubit frequency to electric potential fluctuations from the plunger gate P2. The qubit will undergo dephasing as a result of the energy splitting noise, which will lead to a decay as defined by:
\begin{equation}
\label{eq:pcpmg}
    P(\tau)\propto \exp\left(-\mathcal{X}(\tau)\right) \\
\end{equation}
with $P$ the measured spin-up probability and
\begin{equation}
    \mathcal{X}(\tau) = \int_0^\infty S_{\omega_q}(\omega) \cdot \frac{F_N(\omega, \tau)}{\pi \omega^2}d\omega
\end{equation}
The unitless filter function $F_N$ for the CPMG experiment is defined as follows~\cite{cywinski_how_2008}:
\begin{equation}
    F_N(\omega,\tau)=
    \begin{cases}
    8\sin^4\left(\omega\tau/4\right)\frac{\sin^2\left(N\omega\tau/2\right)}{\cos^2\left(\omega\tau/2\right)},& N \text{ is even}\\
    8\sin^4\left(\omega\tau/4\right)\frac{\cos^2\left(N\omega\tau/2\right)}{\cos^2\left(\omega\tau/2\right)},& N \text{ is odd}
\end{cases}
\end{equation}

As both the strength of the nuclear spin noise and charge noise are expected to depend on $\mathbf{B}$, we fit the data for each $\theta_\mathbf{B}$ independently, fixing $\gamma_\text{Ge-73}=1.48$~MHz/T and keeping $\sigma_\text{Ge-73}$, $S_{0,V}$, and $S_{0,\text{hf}}$ as fit parameters. 

We note that we find that $\sigma_\text{Ge-73}$ is independent of $\theta_\mathbf{B}$ within the experimental range, with an average $\overline{\sigma}_\text{Ge-73}=9$~kHz (see Supp. Fig.~\ref{fig:fig_nuclear_rest}e,f). The finite width of the hyperfine line is mostly reflected in the loss of the coherence for low magnetic fields, when $f_\text{Ge-73}\approx\sigma_\text{Ge-73}$. This can be observed in the data presented in Fig.~\ref{fig:esens2}, as well as when performing the CPMG experiment as a function of the magnetic field strength (see Supp. Fig.~\ref{fig:cpmg_field}). However, we observe this line width to be dependent on the azimuth orientation of the external magnetic field $\phi_\mathbf{B}$ (see Supp. Fig.~\ref{fig:fig_nuclear_rest}), potentially indicative of a quadrupolar origin, which would depend on strain and electric fields and thus be magnetic field orientation dependent.

Increasing the number of refocusing pulses also sharpens the effective band pass filter of the CPMG~\cite{cywinski_how_2008, biercuk_dynamical_2011}, thus enhancing the sensitivity to the nuclear spin precession frequency. As a result, a higher accuracy of $\theta_\mathbf{B}$ is required to align exactly to the hyperfine sweet spot and avoid loss of coherence due to hyperfine interaction with the $^{73}$Ge nuclear spins. This is illustrated in Supp. Fig.~\ref{fig:fig_cpmg_vs_N}, where we measure the CPMG decay as a function of the number of refocusing pulses $N$.

\subsection*{Estimation of the hyperfine coupling constant}
The reconstruction of the hyperfine noise spectrum allows for an estimation of the hyperfine coupling constants for a heavy hole in germanium. From the fit to the data in Fig.~\ref{fig:nuclear_combined} in the main text, we have $S_\text{0,hf}=2.52(4)\text{kHz}^2/\text{Hz}$ for an out-of-plane field and $\sigma_\text{Ge-73}=9.9(11)$~kHz. This equates to an integrated detuning noise of:
\begin{equation}
    \sigma_f=\sqrt{\sqrt{2\pi}S_\text{0,hf}\sigma_\text{Ge-73}}=250~\text{kHz}
\end{equation}
Assuming a Gaussian noise distribution, this corresponds to an expected phase coherence time~\cite{nakajima_coherence_2020} of $T_2^*=1/(\pi\sqrt{2}\sigma_f)=900~\text{ns}$. We can estimate the out-of-plane hyperfine coupling $A_\parallel$ using Eq. 2.65 from Ref.~\cite{philippopoulos_first-principles_2020}:
\begin{equation}
    h^2\sigma_f^2 \approx \frac{1}{4N}g_\text{Ge-73}I(I+1)A_\parallel^2
\end{equation}
Such that:
\begin{align}
\label{eq:Anuc}
    A_\parallel &\approx \sqrt{\frac{4N}{g_\text{Ge-73}I(I+1)}}h\sigma_f
\end{align}
with $g_\text{Ge-73}=0.0776$ the natural abundance of the $^{73}$Ge isotope, $I=9/2$ the $^{73}$Ge nuclear spin and $N$ the number of nuclei the quantum dot wave function overlaps. To estimate N, we consider a cylindrical quantum dot, such that $N=\pi r^2w/v_0$, with $r$ the radius and $w$ the height of the dot, and $v_0=2.3\cdot10^{-29}~\text{m}^3$ the atomic volume of germanium. We can estimate $r$ from the single particle level splitting $\Delta E\approx1.2$~meV as can be obtained from the extend of the PSB readout window, and find $r\approx 35$~nm. This is in good agreement with $r\approx50$~nm as expected from the charging energy $E_\text{C}\approx 2.8$~meV and the capacitance of a disk: $r= e^2/(8\epsilon_rE_c)$. Assuming $r=35$~nm and $w=10$~nm (half of the quantum well width), we then find $N\approx1.7\cdot10^{6}$. Using Eq.~\ref{eq:Anuc}, we estimate the hyperfine coupling constant to be $|A_\parallel|\approx1.9~\mu$eV, which is in good agreement with the theoretical prediction $A_\parallel=-1.1~\mu$eV from Ref.~\cite{philippopoulos_hyperfine_2020}. Similarly from the extracted $S_\text{0,hf}$ for an in-plane $\mathbf{B}$, we estimate an upper bound for the in-plane hyperfine coupling constant $A_\perp<0.1\mu$eV compatible with the predicted $A_\perp=0.02~\mu$eV.

\subsection*{Randomized benchmarking}
To extract the single qubit gate fidelity, we perform randomized benchmarking of the Clifford gate set presented in Supp.~Table~\ref{tab:RB_cliff}. For every randomization, we measure both the projection to $\ket{\uparrow}$ and $\ket{\downarrow}$ and fit the difference to avoid inaccuracies due to the offset of the charge sensor current. The measured current is normalized to the signal obtained from a separate measurement of our $\ket{\uparrow}$ and $\ket{\downarrow}$ states. We fit the data to $P=a\exp(-(1-2F_\text{C})N_\text{C})$, with $F_\text{C}$ the Clifford gate fidelity and $N$ the number of applied Clifford gates. $a$ is an additional scaling parameter we include to account for the reduced visibility we observe when applying a large number of rf pulses. Fixing $a=1$ does not significantly alter the fit as shown by the dashed line in Fig.~\ref{fig:coherence}. In fact, we find $F_\text{g}=99.92$~\% for $T_\text{mc}=20$~mK and $F_\text{g}=99.7$~\% for $T_\text{mc}=1.1$~K when fixing $a=1$. The primitive gate fidelity $F_\text{g}$ can be calculated by accounting for the number of physical gates per Clifford: $0.875$ for this gate set.

\subsection*{Extraction of the $\bm{g}$-tensor sensitivity}
We measure the modulation of the qubit energy splitting $\delta f_\text{Q}$ as the result of a small voltage pulse $\delta V$ on one of the quantum dot gates. The voltage pulse will temporarily shift the qubit resonance frequency, thus inducing an effective phase gate, controlled by the length of the pulse $t_Z$. By incorporating this phase gate within the free evolution of a Hahn echo experiment, we can observe the phase oscillations as a function of $t_Z$, as shown in Fig.~\ref{fig:esens2}c of the main text. From the frequency of these oscillations, we obtain $|\delta f_\text{Q}|$. We confirm that for a small $\delta V$, $|\delta f_\text{Q}|$ is linear in $\delta V$, allowing us to extract the sensitivity $|\partial f_\text{Q}/\partial V_i|$ from a single data point of $\delta V$ (see Supp. Fig.~\ref{fig:lin_meas}). To exclude effects caused by the exchange interaction $J$ between the qubits, we tune $J<1$~MHz using the interdot barrier B12. Furthermore, we tune the device to the point of symmetric exchange in the (1,1) region~\cite{reed_reduced_2016, martins_noise_2016} and apply symmetric pulses in the first and second free evolution period of the Hahn sequence, echoing out effects caused by changes of the double dot detuning. To extract the sign of $\partial f_\text{Q}/\partial V_i$, we measure the qubit resonance frequency for three different gate voltage settings (see Fig.~\ref{fig:esens2}c) for a few selected magnetic field orientations.

Given a g-tensor $\tensor g$ and a g-tensor sensitivity $\partial\tensor{g}/\partial V_i$, $\partial f_\text{Q}/\partial V_i$ only depends on the magnetic field direction $\mathbf{b}$ and on $f_\text{Q}\propto B$:
\begin{equation}
\label{eq: dfdV}
\frac{\delta f_\text{Q}}{\delta V}(\theta_\mathbf{B}, \phi_\mathbf{B}, B) = \frac{\left(\partial\tensor{g}/\partial V_i\cdot\mathbf{b}\right)\cdot\left(\tensor g \cdot \mathbf{b}\right)}{\left(\tensor g \cdot \mathbf b\right)^2}f_\text{Q}(B)
\end{equation}
We extract $\partial\tensor{g}/\partial V_i$ by fitting Eq.~\ref{eq: dfdV} to the data presented in Fig.~\ref{fig:esens_fitting} of the main text, using $\tensor{g}$ as extracted previously and displayed in Fig.~\ref{fig:g_tensor}h. We then calculate the expected $g$-TMR mediated Rabi frequency using
\begin{equation}
\label{eq:dfrdV}
\frac{\delta f_\text{Rabi}}{\delta V}(\theta_\mathbf{B}, \phi_\mathbf{B}, B) = \mu \frac{\left|\left(\partial \tensor{g}/\partial V_i\cdot\mathbf{b}\right)\cross\left(\tensor{g}\cdot\mathbf{b}\right)\right|}{2|\tensor{g}\cdot\mathbf{b}|^2}f_\text{Q}(B)
\end{equation}
with $f_\text{Rabi}$ the Rabi frequency and $\mu$ the signal attenuation for a microwave signal at a frequency of $f_\text{Q}$.

We fit the data to the Eq.~\ref{eq:dfrdV}, with $\mu$ as the only fit parameter. We find a line attenuation of $\mu_\text{P2}=2.1$, $\mu_\text{B2}=2.1$, and $\mu_\text{B12}=2.0$. These value are in good agreement with the attenuation of our experimental setup at $f=225$~MHz as extracted from the broadening of the charge sensor Coulomb peak ($\mu=2.1-2.5$) (see Supp. Fig.~\ref{fig:attenuation}).

\section*{Data availability}
All data underlying this study will be made available in a Zenodo repository.

\section*{Acknowledgements}
We acknowledge the staff of the Binnig and Rohrer Nanotechnology Center for their contributions to the sample fabrication and thank all members of the IBM Research Europe - Zurich spin qubit team for useful discussions. A.F. acknowledges support from the Swiss National Science Foundation through Grant No. 200021 188752.

\section*{Author contributions}
N.W.H. performed the experiments and data analysis, with contributions from G.S.; N.W.H. fabricated the sample with contributions from L.M., M.M., and F.S.; S.P. contributed to the development of the experimental setup. S.W.B. provided the heterostructures. N.W.H. wrote the manuscript with contributions from G.S. and A.F. and input from all authors. N.W.H. and A.F. planned the project.

\section*{Competing Interests}
The authors declare no competing interests.

\section*{Additional information}
Supplementary information is available with this paper.

\clearpage
\title{Supplementary information: Sweet-spot operation of a germanium hole spin qubit with highly anisotropic noise sensitivity}
\maketitle
\renewcommand{\figurename}{Supplementary Figure}
\renewcommand{\tablename}{Supplementary Table}
\setcounter{figure}{0}
\setcounter{page}{1}
\onecolumngrid

\begin{table}
\begin{tabular}{llll}
\toprule
$\partial\tensor{g}/\partial V_i$ &           P2 &           B2 &          B12 \\
\midrule
$\partial\phi/\partial V_i$ (mrad$\cdot$mV$^{-1}$)  &     -0.11(2) &       1.3(2) &      -3.1(1) \\
$\partial\theta/\partial V_i$ (mrad$\cdot$mV$^{-1}$) &    0.0008(5) &     0.005(4) &     0.028(3) \\
$\partial\zeta/\partial V_i$ (mrad$\cdot$mV$^{-1}$)   &      0.04(2) &      -2.3(1) &       2.2(1) \\
$\partial g_{x'}/\partial V_i$ (mV$^{-1}$)        &  0.000181(9) &  -0.00028(7) &  -0.00073(6) \\
$\partial g_{y'}/\partial V_i$ (mV$^{-1}$)        &  0.000507(3) &   0.00037(2) &  -0.00146(2) \\
$\partial g_{z'}/\partial V_i$ (mV$^{-1}$)        &    0.0045(1) &   -0.0001(8) &   -0.0071(7) \\
\bottomrule
\end{tabular}
\caption{\textbf{Description of $\bm{\partial\tensor{g}/\partial V_i}$ of Q2 for a potential applied to gates P2, B2, and B12.}
    Overview of the parameters describing the voltage-induced deformation of the $g$-tensor. Three Euler angles $\partial\zeta/\partial V_i$, $\partial\theta/\partial V_i$, and $\partial\phi/\partial V_i$ describing the rotation, as well as changes to the principle $g$-factors $\partial g_{x'}/\partial V_i$, $\partial g_{y'}/\partial V_i$, and $\partial g_{z'}/\partial V_i$, for Q2.
    }
\label{tab:dg_tensor}
\end{table}

\begin{table}
\begin{tabular}{lll}
\toprule
Data set & $\phi_B$~(deg.) & $\theta_B$~(deg.) \\
\midrule
\tikz\draw[color1,fill=color1] (0,0) circle (.5ex); & 0.0 & 90.03\\
\begin{tikzpicture}
 \node[regular polygon,regular polygon sides=3, color2, draw, shape border rotate=180, fill=color2, inner sep=.25ex] at (0,0){};
\end{tikzpicture}
& 0.0 & 85.50\\
\begin{tikzpicture}
 \node[regular polygon,regular polygon sides=3, color3, draw, fill=color3, inner sep=.25ex] at (0,0){};
\end{tikzpicture}
& 0.0 & 95.63\\
\begin{tikzpicture}
 \node[regular polygon,regular polygon sides=3, color4, draw, rotate=90, fill=color4, inner sep=.25ex] at (0,0){};
\end{tikzpicture}
& -105.0 & 95.63\\
\begin{tikzpicture}
 \node[regular polygon,regular polygon sides=3, color5, draw, rotate=270, fill=color5, inner sep=.25ex] at (0,0){};
\end{tikzpicture}
& -105.0 & 90.00\\
\kern-.25ex\tikz\draw[color6,line width=.5ex] (0ex,0)--(1.25ex,0)(0.625ex,.625ex)--(0.625ex,-.625ex); 
& -105.0 & 89.91\\
\tikz\draw[color7,line width=.5ex] (-.5ex,-.5ex)--(.5ex,.5ex)(0.5ex,-.5ex)--(-0.5ex,.5ex); & 0.0 & 91.13\\
\begin{tikzpicture}
    \node[diamond,draw, fill=color8, color8, inner sep=.35ex] (d) at (0,0) {};
\end{tikzpicture}
& 0.0 & 91.08\\
\bottomrule
\end{tabular}
\caption{\textbf{Magnetic field parameters for data in Fig.~\ref{fig:esens2} of the main text.}
    Magnetic field elevation $\theta_B$ and azimuth $\phi_B$ for the different data sets in Fig.~\ref{fig:esens2} of the main text, as indicated by the coloured markers.
    }
\label{tab:field_angle}
\end{table}

\begin{table}
\begin{tabular}{|l|}
\hline
Clifford gates \\ \hline
I\\ \hline
-Z, X/2, X/2\\ \hline
-Z, X/2\\ \hline
-Z, X/2, -Z\\ \hline
-Z, X/2, -Z/2\\ \hline
-Z, X/2, Z/2\\ \hline
-Z/2\\ \hline
-Z/2, X/2\\ \hline
-Z/2, X/2, -Z\\ \hline
-Z/2, X/2, -Z/2\\ \hline
-Z/2, X/2, Z/2\\ \hline
X/2, X/2\\ \hline
X/2, X/2, Z/2\\ \hline
X/2\\ \hline
X/2, -Z\\ \hline
X/2, -Z/2\\ \hline
X/2, Z/2\\ \hline
Z\\ \hline
Z/2\\ \hline
Z/2, X/2, X/2\\ \hline
Z/2, X/2\\ \hline
Z/2, X/2, -Z\\ \hline
Z/2, X/2, -Z/2\\ \hline
Z/2, X/2, Z/2\\ \hline
\end{tabular}
\caption{\textbf{Clifford set used for randomized benchmarking.}
    The Clifford set used for the randomized benchmarking consists out of $X_{\pi/2}$ pulses, and virtual $Z$ gates. The average number of physical gates per Clifford equals 0.875.
    }
\label{tab:RB_cliff}
\end{table}

\FloatBarrier
\begin{figure*}[htp]
\includegraphics{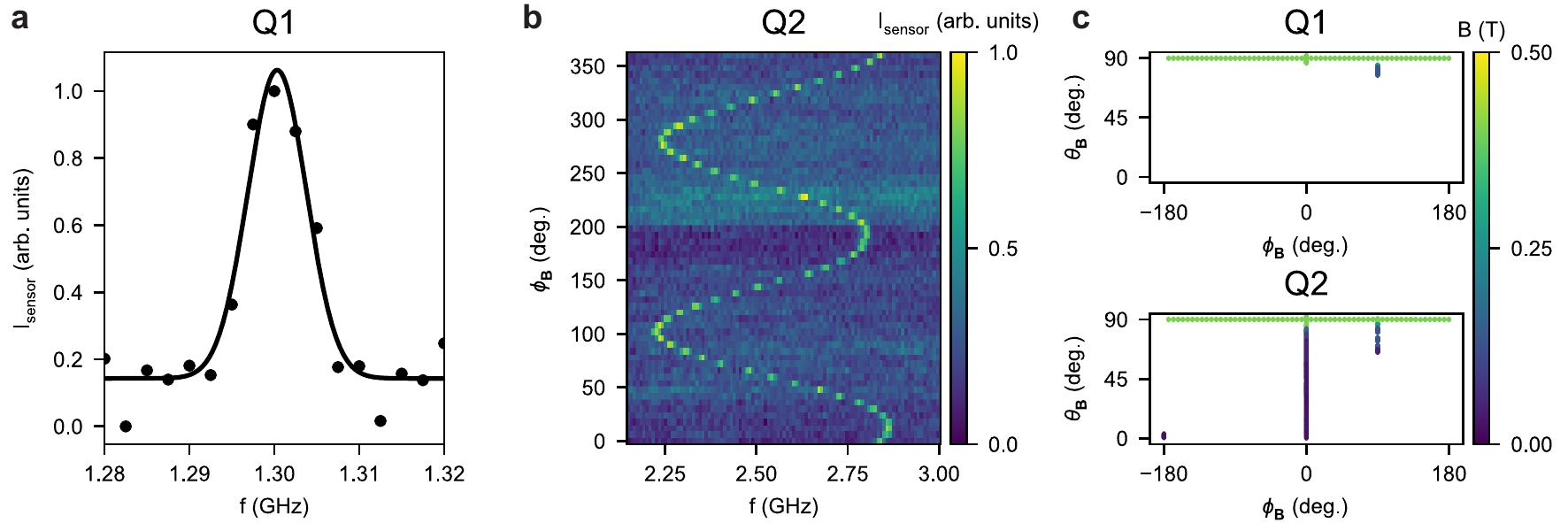}
    \caption{\textbf{Exemplary EDSR resonance traces used to fit the qubit $g$-tensors.}
    \textbf{a}, Normalized charge sensor current, as a function of the microwave frequency $f$, showing spin resonance for Q1. We apply a frequency chirp to assure full spin inversion.
    \textbf{b}, Normalized charge sensor current, as a function of the microwave frequency $f$ and magnetic field azimuth $\phi_\mathbf{B}$.
    \textbf{c}, Illustration of all magnetic field orientations at which $g^*$ was measured to fit $\tensor{g}$, for Q1 (top) and Q2 (bottom). Each of the 402 (935) markers corresponds to a single resonance measurement for Q1 (Q2) as in panel \textbf{a}.
    }
\label{fig:edsr_spectra}
\end{figure*}

\begin{figure*}[htp]
\includegraphics{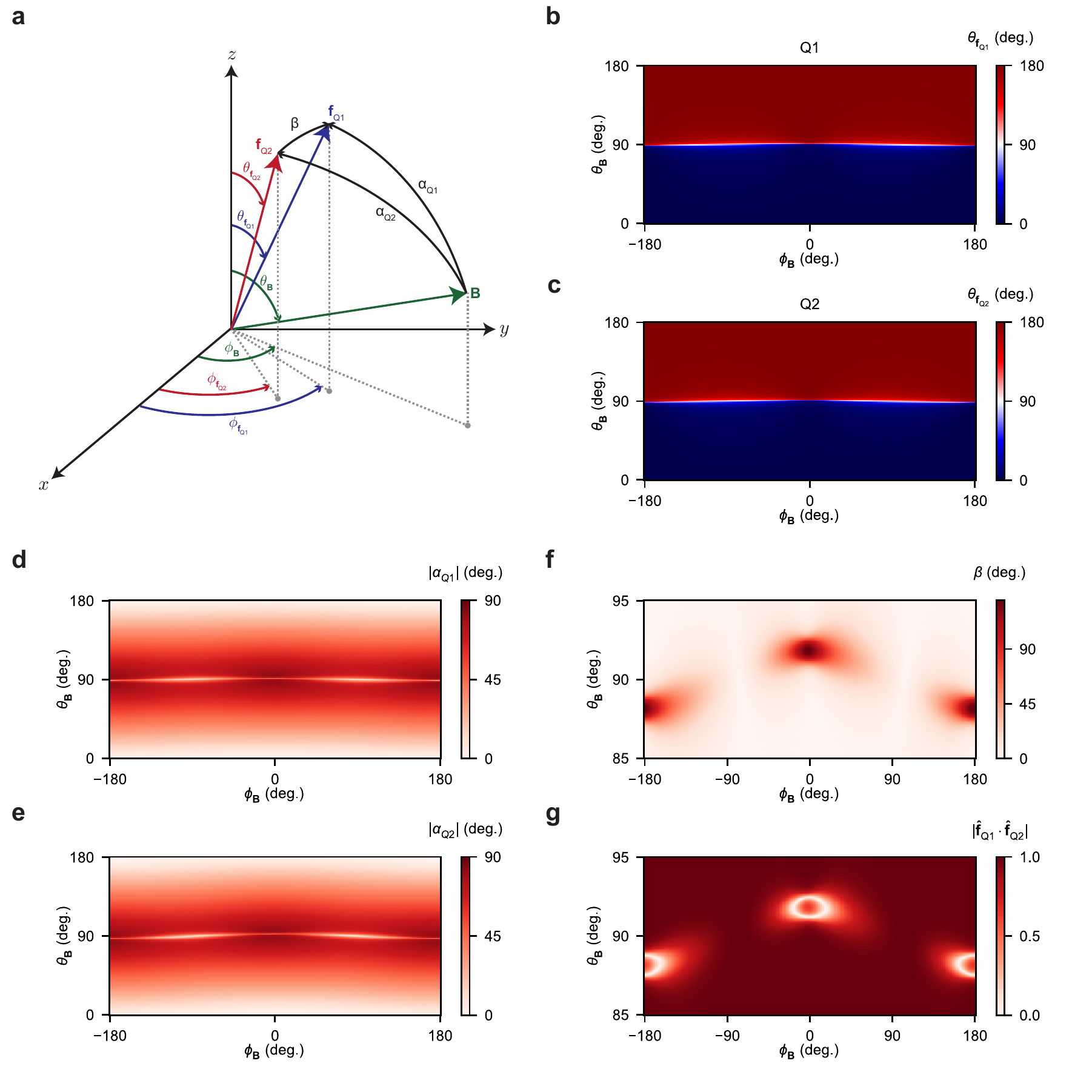}
    \caption{\textbf{Projections of the fitted $\bm{g}$-tensors.} From the experimentally extracted the qubit $g$-tensors, we calculate the qubit quantisation axis $h\mathbf{f_\text{Q}}=\mu_\text{B}\tensor{g}\mathbf{B}$ as a function of $\mathbf{B}$.
    \textbf{a}, Diagram illustrating the relevant angles. $\theta_\mathbf{B}$, and $\phi_\mathbf{B}$ are the elevation and azimuth angle of the applied magnetic field respectively. $\theta_{\mathbf{f_{\text{Q}i}}}$, and $\phi_{\mathbf{f_{\text{Q}i}}}$ are the elevation and azimuth angle of the resulting Larmor vector of Q$i$. $\alpha_{Qi}$ is the angle between the applied magnetic field and the resulting Larmor vector of Q$i$ and $\beta$ is the angle between the two quantisation axes of the two qubits.
    \textbf{b},\textbf{c}, Elevation angle of the Larmor vector of Q1 (\textbf{b}) and Q2 (\textbf{c}), as a function of the orientation of the magnetic field. As a result of the large anisotropy of the $g$-tensor, the quantisation axis of the qubit rapidly flips from $\mathbf{f_\mathbf{\text{Q}}}\parallel z$ to $\mathbf{f_\mathbf{\text{Q}}}\parallel -z$ as the magnetic field crosses the equator of the $g$-tensor.
    \textbf{d},\textbf{e}, The absolute angle between the qubit quantisation axis and applied magnetic field direction $\abs{\alpha_{Qi}}$ for Q1 (\textbf{d}) and Q2 (\textbf{e}).
    \textbf{f}, Angle between the qubit quantisation axes $\beta$ as a function of the magnetic field orientation. Near the principle axis directions of the qubit $g$-tensors, a large misalignment between the two quantisation axes can be observed.
    \textbf{g}, Colour plot of $\abs{\hat{\mathbf{f}}_{\mathbf{Q1}}\cdot\hat{\mathbf{f}}_{\mathbf{Q2}}}$, illustrating the orthogonality of the two qubit quantisation axes. For the ring-shaped regions where this quantity equals $0$, around the $x$ principle axis of the $g$-tensors, the qubit quantisation axes of Q1 and Q2 are perpendicular to each other.}
\label{fig:projection}
\end{figure*}

\begin{figure*}[htp]
\includegraphics{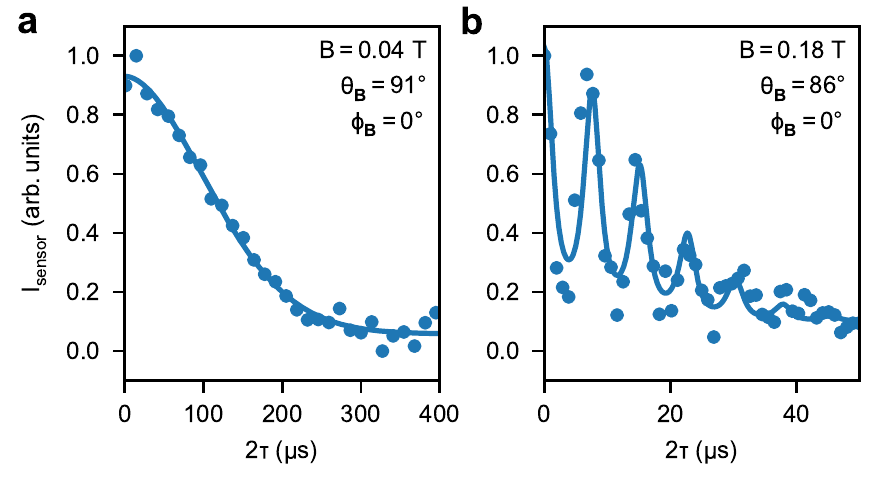}
\caption{\textbf{Hahn echo decay of Q2.} 
    \textbf{a}, Hahn echo decay of Q2, for the magnetic field aligned with the hyperfine sweet plane. Solid line is a fit to the data used to extract $T_2^\text{H}$.
    \textbf{b}, Hahn echo decay of Q2, for the magnetic field aligned away from the hyperfine sweet plane. A collapse-and-revival structure can be observed due to the interaction of the qubit with the precession of the $^{73}$Ge nuclear spins. Solid line is a fit to the data used to extract $T_2^\text{H}$.
    }
\label{fig:hahn_example}
\end{figure*}

\begin{figure*}[htp]
\includegraphics{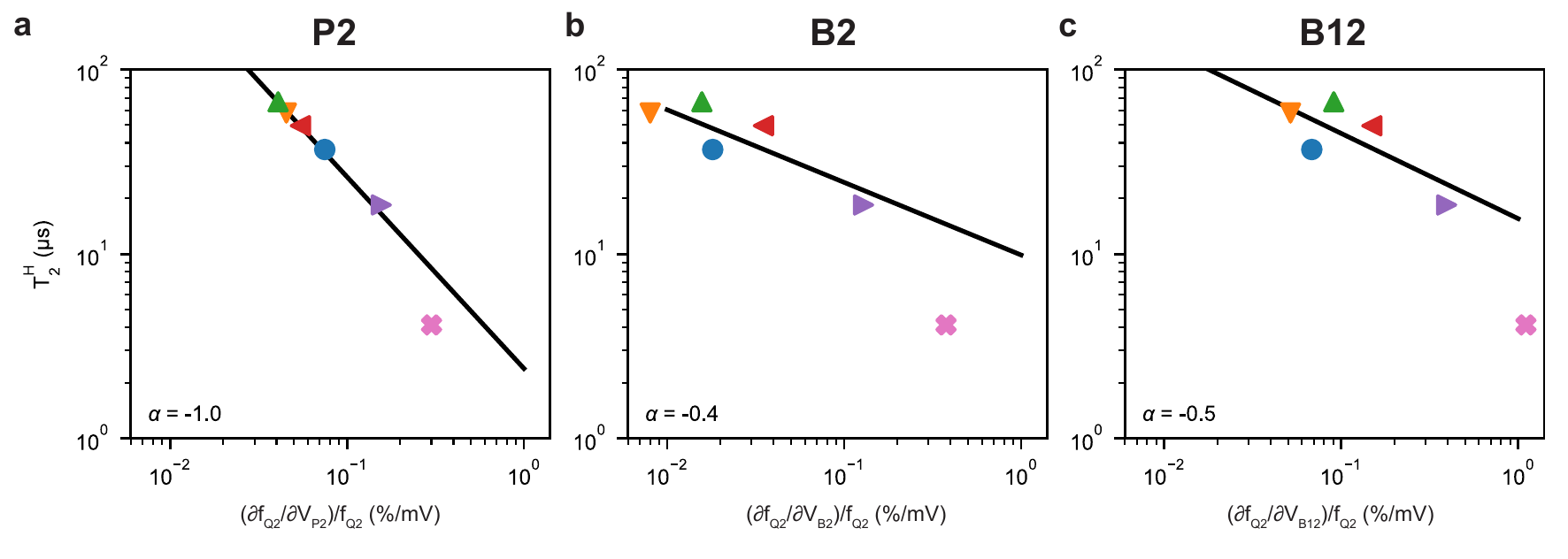}
\caption{\textbf{Q2 coherence as a function of the different gate voltage sensitivities}
    \textbf{a-c}, Observed Hahn echo coherence time (data from Fig.~\ref{fig:esens2}g in the main text), as a function of the relative qubit frequency sensitivity $(\partial f_\text{Q2}/\partial V_i)/f_\text{Q2}$ to a voltage fluctuation on gate P2 (\textbf{a}), B2 (\textbf{a}), or B12 (\textbf{a}). Voltage sensitivities are obtained from the data for gate P2 (Fig.~\ref{fig:esens2}e) from the fits of $\delta \tensor{g}$ presented in Fig.~\ref{fig:esens_fitting} of the main text, as well as Fig.~\ref{fig:fitprojections} of this Supplementary Information for gates B2 and B12. Black lines are fits to a power law $T_2^H=a\exp(((\partial f_\text{Q2}/\partial V_i)/f_\text{Q2})^\alpha)$, with $a$ a scaling factor and $\alpha$ the exponent. We find the correct trend with $\alpha=-1$ (see Methods) for a voltage noise effectively coming from the plunger gate of the qubit, indicating the dominant charge traps are located directly above our qubit for most magnetic field orientations. For very specific field orientations where the qubit sensitivity to noise from other directions is particularly strong, such as for the pink data point, differently located charge traps could limit $T_2^H$.
    }
\label{fig:gate_comparison}
\end{figure*}

\begin{figure*}[htp]
\includegraphics{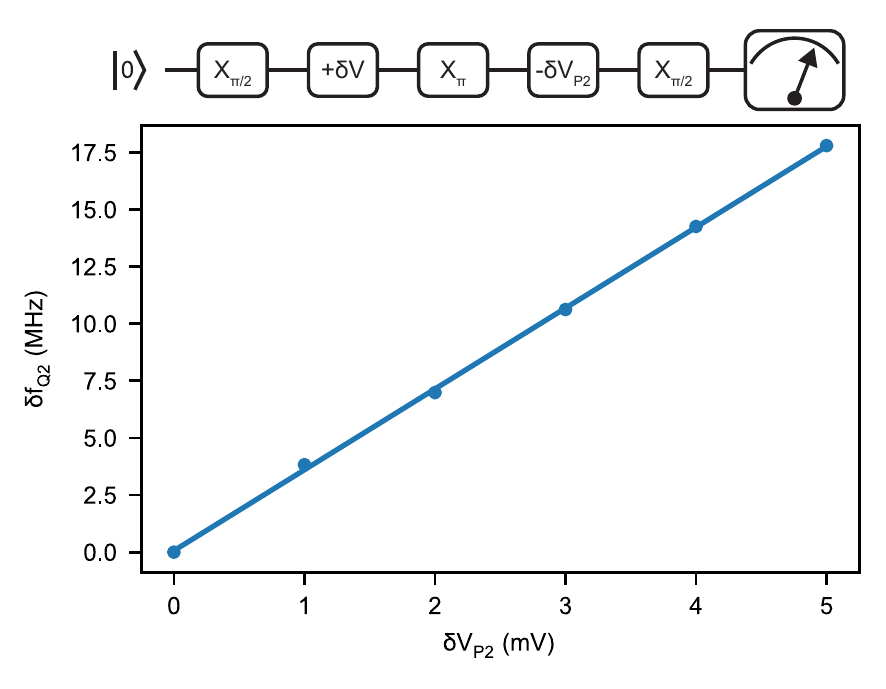}
\caption{\textbf{Linearity of the qubit frequency shift.} 
    We apply a Hahn echo sequence to the qubit, where a positive (negative) voltage pulse $\delta V_\text{P2}$ is applied during the first (second) free evolution, as illustrated in the schematic. As we vary the depth of the voltage pulse, we observe a linear shift of the qubit frequency, which allows us to extract $\partial f_\text{Q2}/\partial V_\text{P2}$ by evaluating the frequency shift for a single point of $\delta V_\text{P2}$.
    }
\label{fig:lin_meas}
\end{figure*}

\begin{figure*}[htp]
\includegraphics{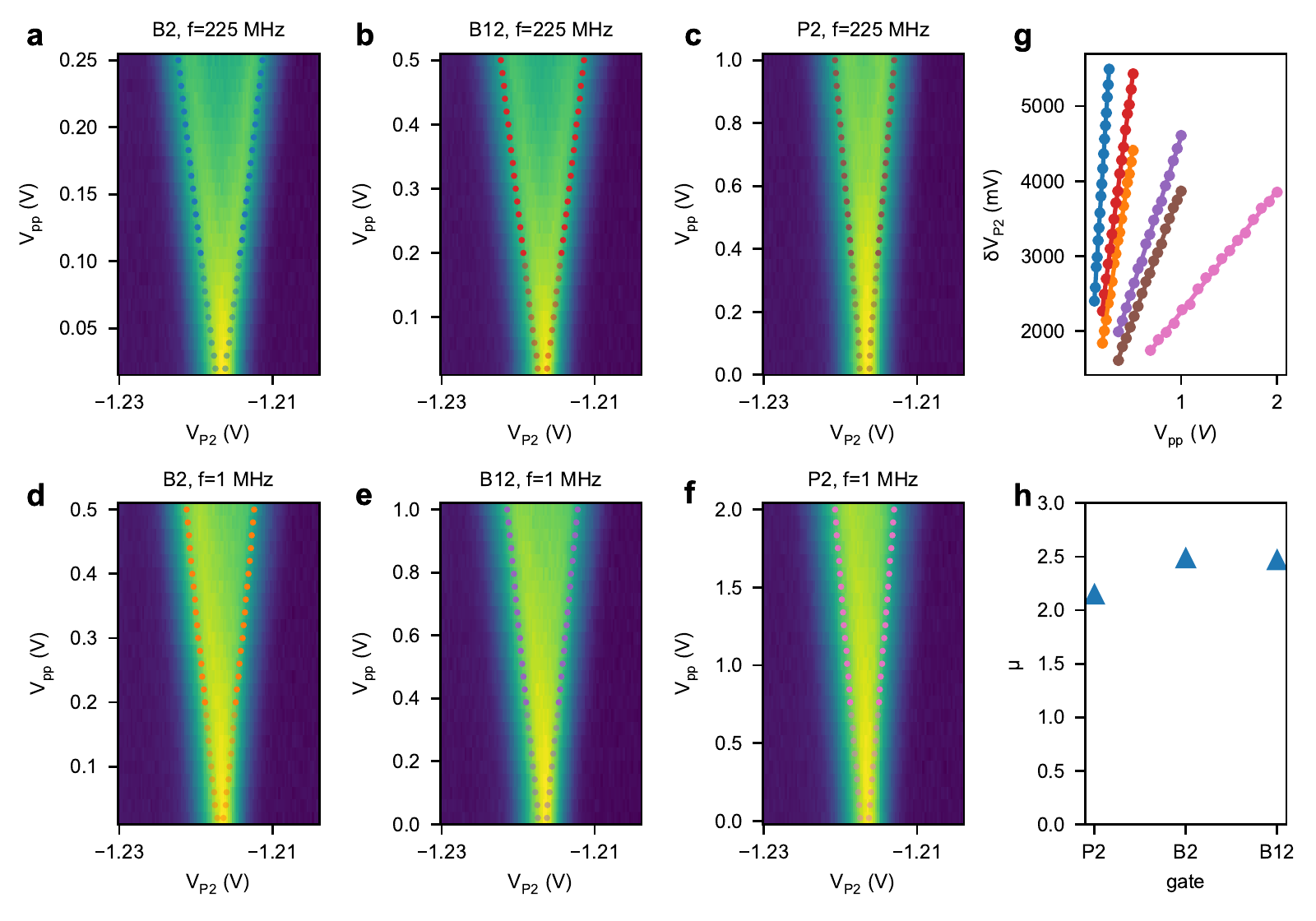}
\caption{\textbf{Attenuation of the microwave pulses applied to different gates.} 
    \textbf{a}-\textbf{c}, Charge sensor Coulomb peak as a result of a continuous wave (CW) microwave excitation with a frequency of $f_\text{CW}=250$~MHz and a peak-to-peak amplitude of $V_\text{pp}$, applied to gate B2 (\textbf{a}), B12 (\textbf{b}), and P2 (\textbf{c}). Solid markers indicate the regime where the splitting $\delta V_\text{pp}$>$\sigma$ with $\sigma$ the peak width.
    \textbf{d}-\textbf{f}, Charge sensor Coulomb peak as a result of a continuous wave microwave excitation with a frequency of $f_\text{CW}=1$~MHz and a peak-to-peak amplitude of $V_\text{pp}$, applied to gate B2 (\textbf{a}), B12 (\textbf{b}), and P2 (\textbf{c}). Solid markers indicate the regime where the splitting $\delta V_\text{pp}$>$\sigma$ with $\sigma$ the Coulomb peak width.
    \textbf{g} Splitting of the Coulomb peak $\delta V_\text{pp}$, as extracted by fitting each line to  the expected shape of a sine wave broadened Lorentzian, obtained by analytically evaluating the convolution between a Lorentzian and the probability distribution of a sine. We extract the relative slope $\alpha$ by fitting the data indicated by the solid markers in panels~\textbf{a}-\textbf{f} to $\delta V_\text{P2}=\alpha V_\text{pp}$. Colour of the markers correspond to the coloured markers as used in panels~\textbf{a}-\textbf{f}.
    \textbf{h} The ratio between the attenuation at $f_\text{CW}=250$~MHz and $f_\text{CW}=1$~MHz for the three different gates. We find attenuation ratios of $\mu_\text{P2}=2.1$, $\mu_\text{B12}=2.5$, and $\mu_\text{B2}=2.5$.
    }
\label{fig:attenuation}
\end{figure*}

\begin{figure*}[htp]
\includegraphics{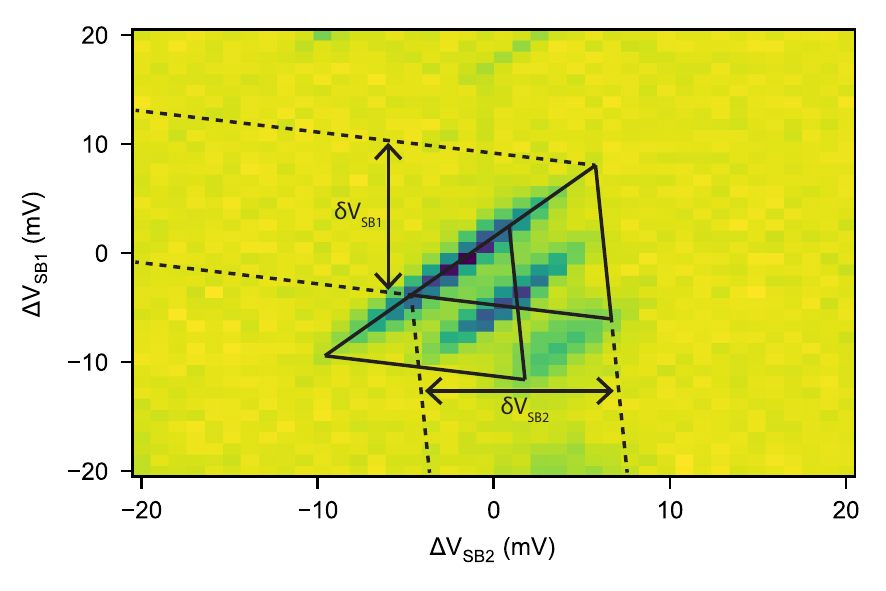}
\caption{\textbf{Lever arm extraction of the plunger gates.} 
    We extract the lever arm of the plunger gates by tuning the charge sensor into a transport double quantum dot underneath gates SB1 and SB2, which are lithographically identical to qubit gates P1 and P2. We measure the bias triangles at $V_\text{SD}=1$~mV and find an average plunger lever arm of $\alpha_\text{P}=7.4(8)$~\%, with the uncertainty as obtained from the scatter of $\alpha$ for gates SB1 and SB2.
    }
\label{fig:leverarm}
\end{figure*}

\begin{figure*}[htp]
\includegraphics{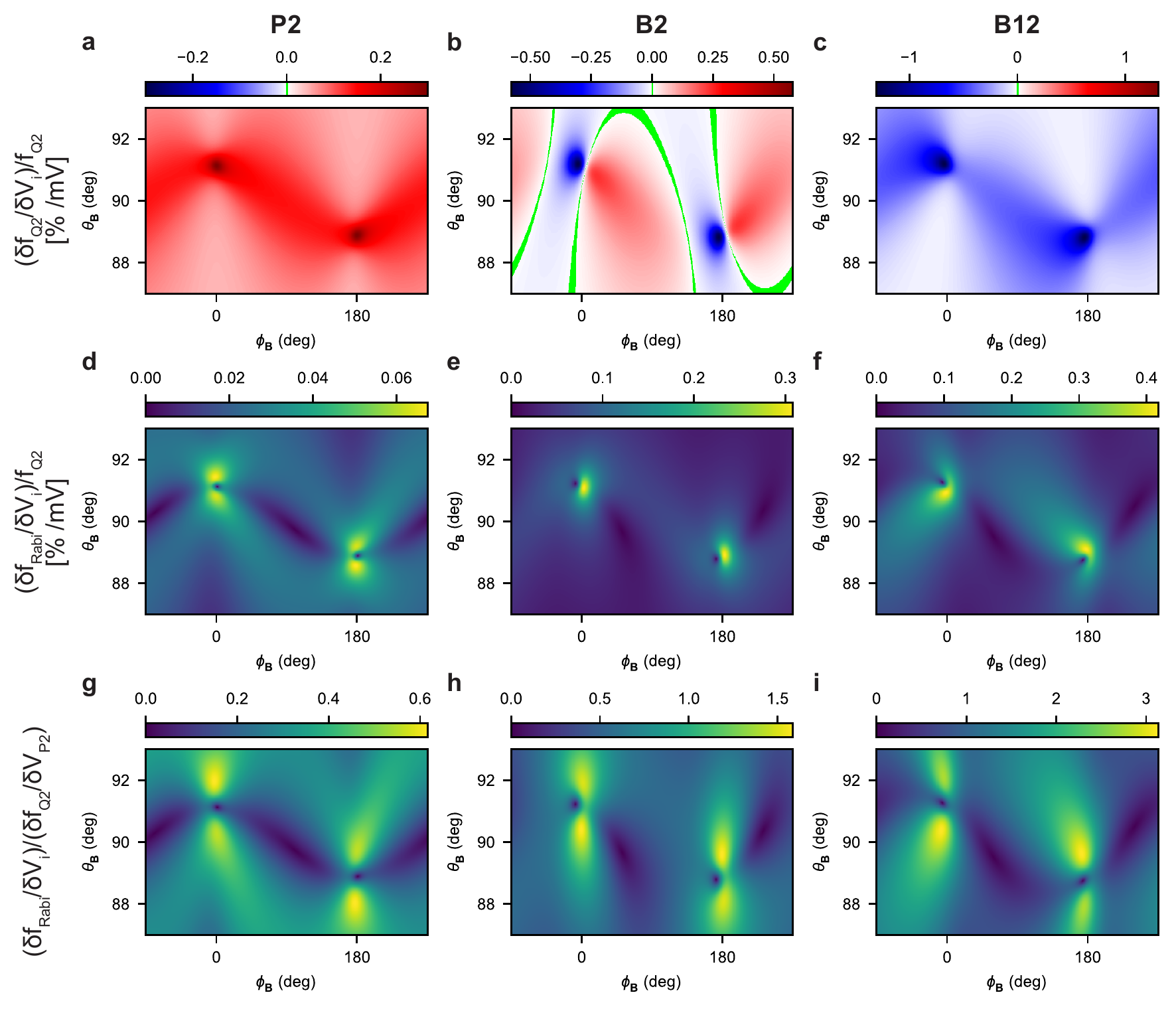}
\caption{\textbf{Longitudinal and transverse components of the fitted $\bm{\partial\tensor{g}/\partial V_i}$.} 
    \textbf{a}-\textbf{c}. Using the fitted $\partial\tensor{g}/\partial V_i$, as detailed in Fig.~\ref{fig:esens_fitting} of the main text and Supp. Table~\ref{tab:dg_tensor} , we plot the expected normalized resonance frequency fluctuation of Q2 as a result of a voltage fluctuation on gate P2 (\textbf{a}), B2 (\textbf{b}), and B12 (\textbf{c}) for different magnetic field orientations. Zero crossings are marked in green, to indicate the presence of a true sweet spot.
    \textbf{d}-\textbf{f}. Expected normalized Rabi frequency fluctuation of Q2 as a result of a drive excitation with amplitude $V_i$ on gate P2 (\textbf{d}), B2 (\textbf{e}), and B12 (\textbf{f}) for different magnetic field orientations.
    \textbf{g}-\textbf{i}. Expected ratio of the transverse and longitudinal components $(\partial f_\text{Rabi}/\partial V_i)/(\partial f_\text{Q2}/\partial V_\text{P2})$ as a result of a drive amplitude on gate P2 (\textbf{g}), B2 (\textbf{h}), and B12 (\textbf{i}) for different magnetic field orientations. We assume the noise to couple in predominantly as if it is applied to the plunger gate P2.
    }
\label{fig:fitprojections}
\end{figure*}

\begin{figure*}[htp]
\includegraphics{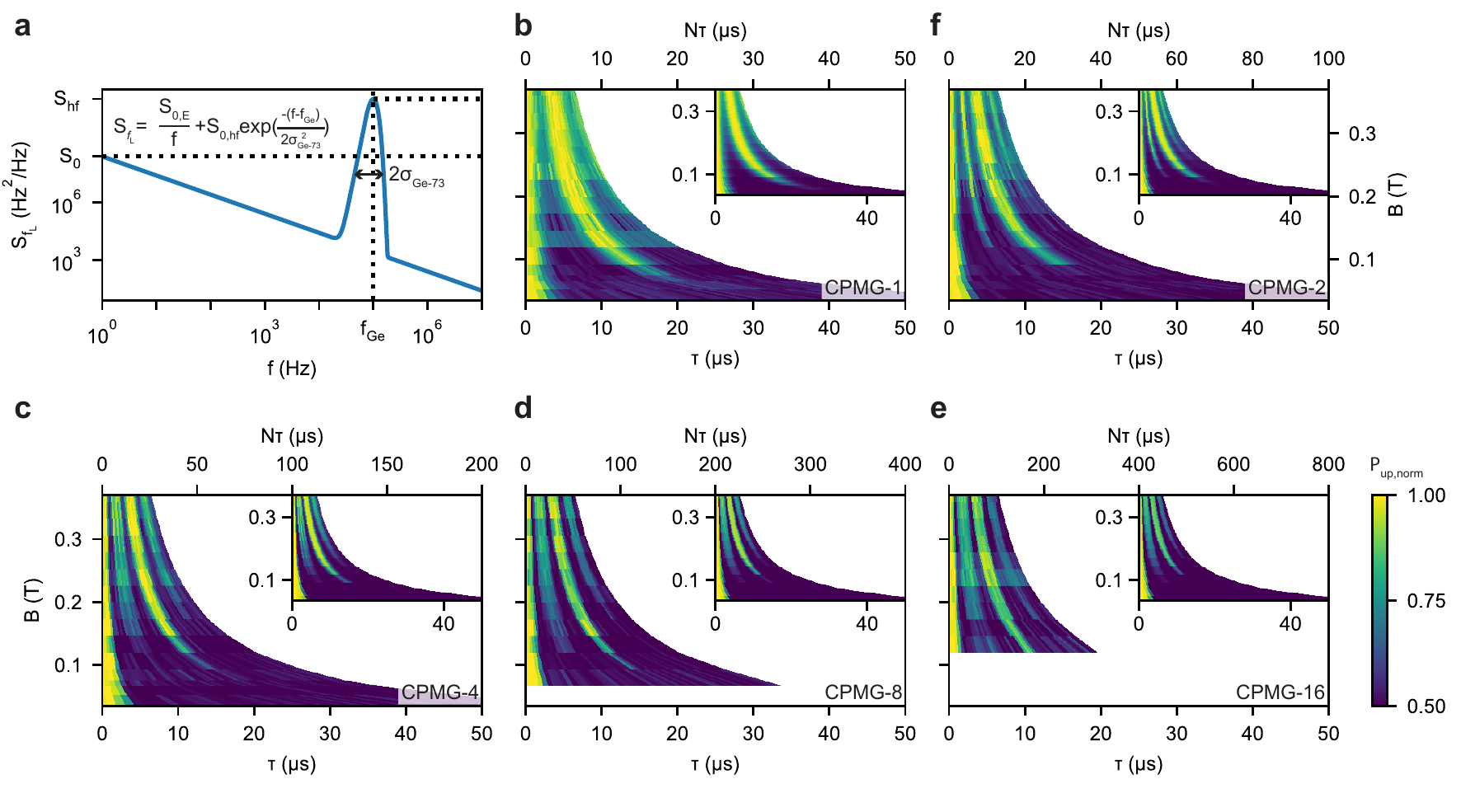}
\caption{\textbf{CPMG dynamical decoupling as a function of magnetic field strength.} 
    \textbf{a}, Schematic illustrating the power spectrum of noise assumed for the fitting. The spectrum consists of a $1/f$ charge noise component with a power $S_{0,\text{E}}$ at 1~Hz, plus a Gaussian spectral line at $f_\text{Ge-73}$ with width $\sigma_\text{Ge-73}$ and power $S_{0,\text{hf}}$.
    \textbf{b}-\textbf{f}, Normalized charge sensor signal for a CPMG sequence with 1 (\textbf{b}), 2 (\textbf{c}), 4 (\textbf{d}), 8 (\textbf{e}), and 16 (\textbf{f}) decoupling pulses respectively, as a function of the spacing between two subsequent decoupling pulses $\tau$ and $B$. $N\tau$ is the total evolution time. $\phi_\mathbf{B}=0\degree$ and $\theta_\mathbf{B}=90\degree$. The inset displays the fit to the data from which we extract $\gamma_\text{Ge-73}=1.48$~MHz/T and $\sigma_\text{Ge-73}=17$~kHz. The extracted charge-induced noise spectrum $S_{0,\text{E}}(B)$ is plotted in Supp.~Fig.~\ref{fig:fig_nuclear_rest}f.
    }
\label{fig:cpmg_field}
\end{figure*}

\begin{figure*}[htp]
\includegraphics{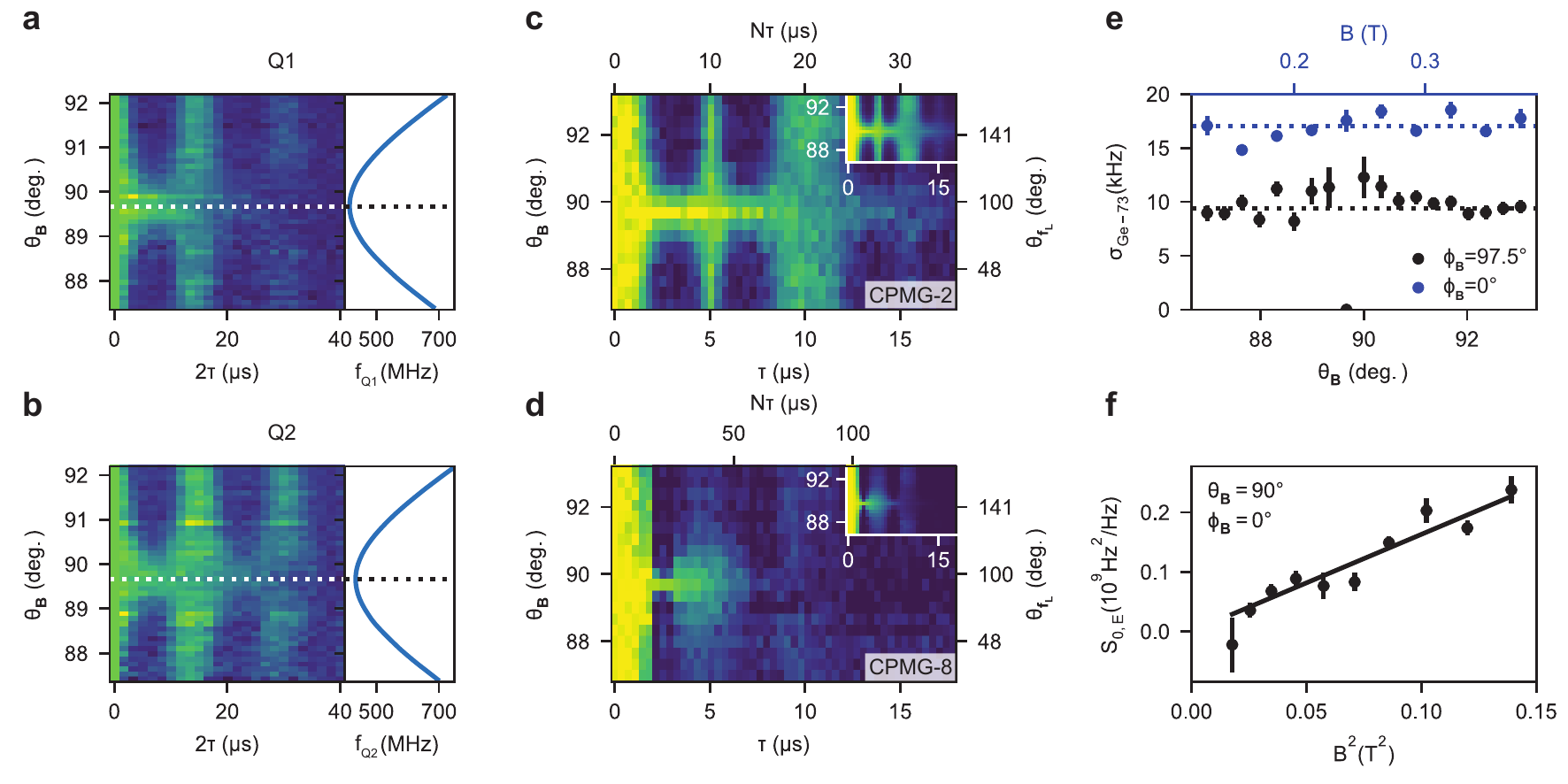}
\caption{\textbf{Hyperfine interaction between the qubits and the $^{\bm{73}}$Ge nuclear spins.}
    \textbf{a},\textbf{b}, Normalized charge sensor signal for a Hahn echo experiment as a function of the total free evolution time $2\tau$ (colour map in left subpanel) and the qubit frequency $f_{\text{Q}i}$ (plot in right subpanel) as a function of the elevation angle of the magnetic field for Q1 (\textbf{a}) and Q2 (\textbf{b}). $B=89$~mT is kept constant throughout the measurement. For any $\theta_\mathbf{B}$ away from the $x'y'$ plane of the $g$-tensor, we observe the collapse-and-revival characteristic for a well-defined spectral component acting on the qubit. Due to the small tilt between the two qubit $g$-tensors, a common hyperfine sweet spot is defined by the intersection of two differently tiled ellipsoids in the lab frame, at $\phi_\mathbf{B}=97.5\degree$, $\theta_\mathbf{B}=89.7\degree$. Selecting this magnetic field orientation allows to operate both qubits in their respective hyperfine sweet planes simultaneously. Here, we find $T_\text{2,Q1}^H=8.1(2)~\mu$s and $T_\text{2,Q2}^H=11.5(6)~\mu$s for $\mathbf{B}$ as specified.
    \textbf{c},\textbf{d}, Normalized charge sensor signal for a CPMG sequence with respectively 2 (\textbf{c}), and 8 (\textbf{d}) decoupling pulses, as a function of the spacing between two subsequent decoupling pulses $\tau$ and $\theta_\mathbf{B}$. $N\tau$ is the total evolution time. The magnetic field strength is $B=133$~mT and the azimuth angle is $\phi_\mathbf{B}=97.5\degree$. The inset displays the fit to the data. These data complement the dataset displayed in Fig.~\ref{fig:nuclear_combined}c,d of the main text.
    \textbf{e}, Width of the hyperfine line $\sigma_\text{Ge-73}$ as extracted from the CPMG coherence as a function of $B$ (Supp. Fig.~\ref{fig:cpmg_field}, $\phi_\mathbf{B}=0\degree$, blue markers), and as extracted from the CPMG coherence as a function of $\theta_\mathbf{B}$ (Fig.~\ref{fig:nuclear_combined}c,d of the main text, $\phi_\mathbf{B}=97.5\degree$, black markers). We find $\sigma_\text{Ge-73}$ to be independent of $\theta_\mathbf{B}$ in the experimental range, but observe a significant difference for the different azimuth directions. Dashed lines correspond to the average $\overline{\sigma}_\text{Ge-73}=9.4$~kHz for $\phi_\mathbf{B}=97.5\degree$ and ${\overline{\sigma}_\text{Ge-73}}=17$~kHz for $\phi_\mathbf{B}=0\degree$.
    \textbf{f}, The intensity of the $1/f$ component of the spectrum at $1$~Hz extracted from the fit to the data in Supp. Fig.~\ref{fig:cpmg_field}. We find $S_{0,\text{E}}\propto B^2\propto f_\text{Q2}^2$, as expected for charge noise with a $1/f$ spectrum.
    }
\label{fig:fig_nuclear_rest}
\end{figure*}

\begin{figure*}[htp]
\includegraphics{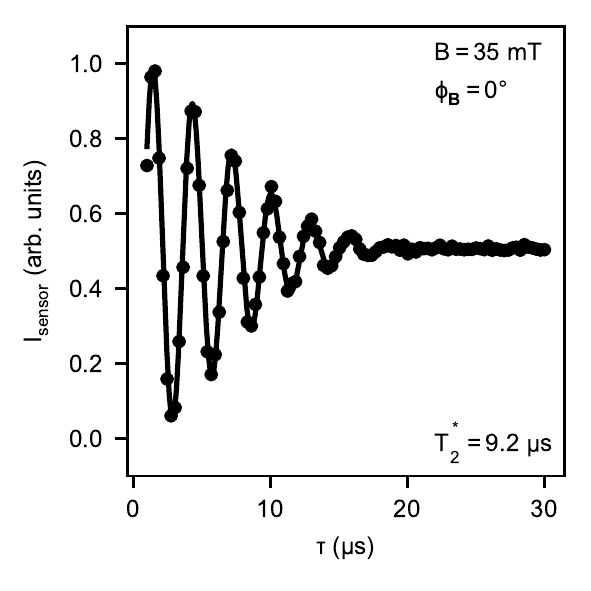}
\caption{\textbf{Extended Ramsey measurement at $\bm{B=35}$~mT.} 
    Normalized charge sensor signal as a function of the waiting time $\tau$ for a Ramsey experiment. The data constitute of an average of 750 traces, for a total integration time of over 12 hours and we find a coherence time of $T_2^*=9.2~\mu$s.
    }
\label{fig:ramsey_20mT}
\end{figure*}

\begin{figure*}[htp]
\includegraphics{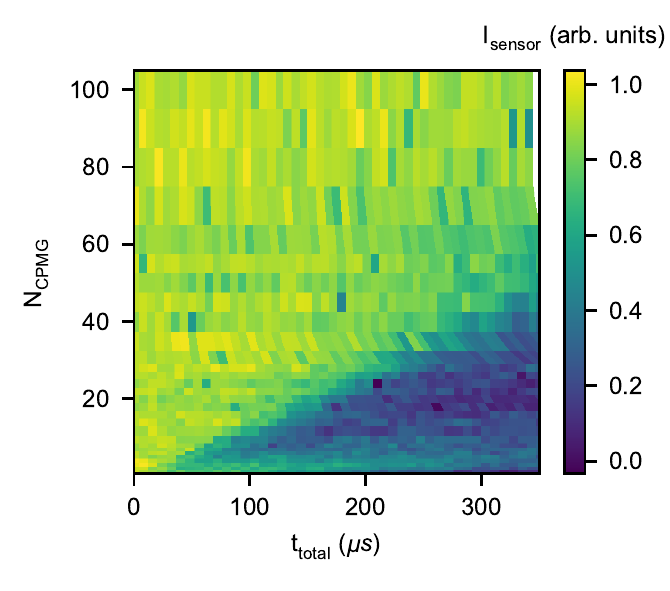}
\caption{\textbf{CPMG dynamical decoupling as a function of the number of decoupling pulses.} 
    Charge sensor signal as a function of the total evolution time and number of decoupling pulses of a CPMG sequence. The magnetic field is aligned to the hyperfine sweet spot. For low $N_\text{CPMG}$ we observe an expected Gaussian decay due to charge noise, but for increasing $N$ the sensitivity to the nuclear noise is enhanced and we observe a sharp coherence collapse caused by the hyperfine interaction. 
    }
\label{fig:fig_cpmg_vs_N}
\end{figure*}

\begin{figure*}[htp]
\includegraphics{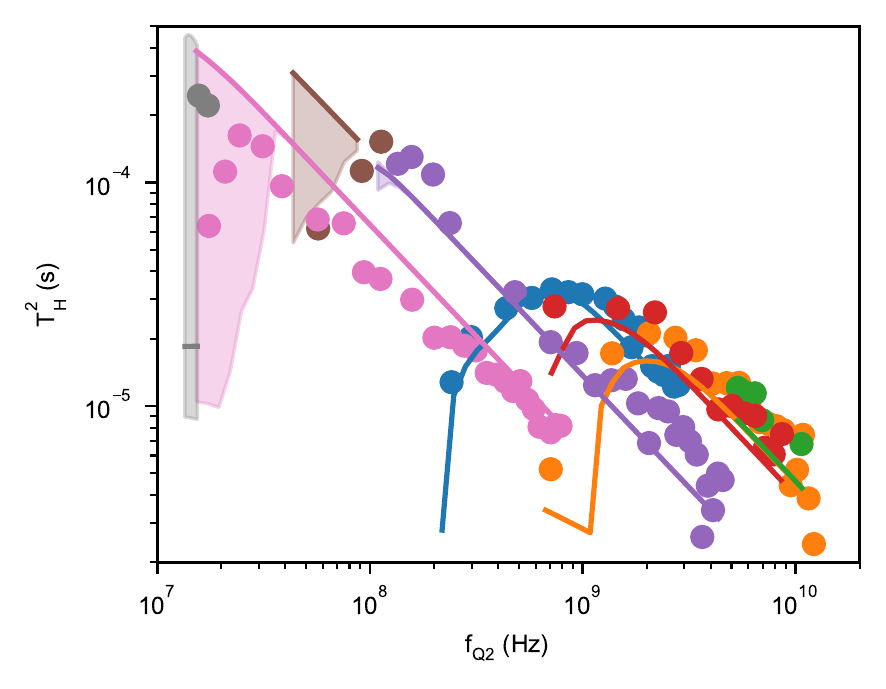}
\caption{\textbf{Expected qubit coherence from extracted noise parameters.}
    We simulate the expected CPMG-1 decay using the filter formalism as detailed in the Methods section, using the noise power parameters as extracted from the CPMG experiment displayed in Fig.~\ref{fig:nuclear_combined} of the main text:  $\sqrt{S_V}=24.7~\mu V/\sqrt{\text{Hz}}$, $\gamma_\text{Ge-73}=1.48$~MHz/T, $S_{0,\text{hf}}=2.5\cdot 10^{12}\cos^2(\theta_{f_\text{Q2}})$, $\sigma_\text{Ge-73}=17$~kHz for $\phi_\mathbf{B}=0\degree$, and $\sigma_\text{Ge-73}=9$~kHz for $\phi_\mathbf{B}=-105\degree$. We then fit the simulated decay using the same procedure as used in Fig.~\ref{fig:esens2}f of the main text to extract the envelope $T_2^\text{H}$. Markers indicate the experimental results and the solid lines correspond to the envelope decay time as predicted by the model. The excellent agreement between the simulation and data, without the need for any fitting parameters, confirms our understanding of the system. The $1/f_\text{Q2}$ decay of the envelope coherence can be explained by an effective voltage noise on plunger gate P2, while the low-$B$ drop-off is caused by the finite spread of the $^{73}$Ge precession frequencies. The shaded area indicates the uncertainty in $T^2_\text{H}$, given an uncertainty of $\pm20~\mu$T in the $z$-component of the magnetic field. For very small $B$, this yields a significant uncertainty in $\theta_\mathbf{B}$, thus complicating an accurate prediction of $T^2_\text{H}$.
    }
\label{fig:fig_fitted_coherence}
\end{figure*}
\end{document}